\documentclass[article,twoside,twocolumn,english,secnumarabic,
footinbib,tightenlines,nobibnotes,aps,prx,unsortedaddress,showpacs,superscriptaddress]{revtex4-2}

\usepackage[range-units=single,separate-uncertainty]{siunitx}
\usepackage[color links=true,linkcolor=blue,citecolor=blue,urlcolor=blue]{hyperref}
\usepackage{relsize}

\let\tensor\relax

\usepackage{LukasTeX2}
\usepackage{graphicx}

\renewcommand{\selectlanguage}[1]{}

\newcolumntype{P}[1]{>{\centering\arraybackslash}p{#1}}

\hypersetup{
    colorlinks,
    linkcolor={blue},
    citecolor={blue},
    urlcolor={blue}
}
\newcolumntype{P}[1]{>{\centering\arraybackslash}p{#1}}

\newcommand{\figletter}[1]{{\textcolor{blue}{#1}}}
\newcommand{\appendixref}[1]{{\textcolor{blue} {\appendixname\ \ref{#1}}}}

\renewcommand{\i}{\text{i}}

\renewcommand{\a}{{\alpha }}
\renewcommand{\b}{{\beta }}
\renewcommand{\d}{{\delta }}
\newcommand{\n}{n}
\newcommand{\g}{{\gamma }}
\renewcommand{\k}{{\kappa }}
\renewcommand{\l}{{\lambda }}
\renewcommand{\o}{\omega}
\newcommand{\q}{q}
\newcommand{\s}{\sigma}
\renewcommand{\t}{\tau}
\newcommand{\eps}{\varepsilon}
\newcommand{\teps}{\tilde\varepsilon}
\newcommand{\tv}{\tilde v}
\newcommand{\pu}{{u'}}
\newcommand{\pv}{{v'}}
\newcommand{\D}{\Delta}
\newcommand{\tot}{{\text{tot}}}

\newcommand{\spi}{\sqrt{\pi}}
\newcommand{\hX}{{\hat X}}
\newcommand{\hY}{{\hat Y}}

\newcommand{\hH}{{\hat H}}
\newcommand{\hs}{{\hat\sigma}}

\renewcommand{\S}{\text{S}}
\newcommand{\erfi}{\text{erfi}}
\newcommand{\erf}{\text{erf}}
\newcommand{\B}{\text{B}}
\newcommand{\RR}{\text{R}}
\newcommand{\Q}{\text{Q}}

\newcommand{\p}{\text{p}}
\newcommand{\E}{\mathbb{E}}
\renewcommand{\P}{\mathbb{P}}
\newcommand{\N}{\mathcal{N}}
\renewcommand{\O}{\mathcal{O}}
\newcommand{\G}{\mathcal{G}}
\newcommand{\tG}{\mathcal{\tilde G}}
\newcommand{\R}{\mathcal{R}}

\newcommand\ETH{{\smaller ETH}}
\newcommand\RMT{{\smaller RMT}}

\newcommand\initial{\text{in}}

\def\ie{{i.e.},\ }
\def\eg{{e.g.}\ }

\def\cf{{cf.}\ }
\def\appendixname{Appendix}

\def\Xint#1{\mathchoice
   {\XXint\displaystyle\textstyle{#1}}%
   {\XXint\textstyle\scriptstyle{#1}}%
   {\XXint\scriptstyle\scriptscriptstyle{#1}}%
   {\XXint\scriptscriptstyle\scriptscriptstyle{#1}}%
   \!\int}
\def\XXint#1#2#3{{\setbox0=\hbox{$#1{#2#3}{\int}$}
     \vcenter{\hbox{$#2#3$}}\kern-.5\wd0}}

\def\dashint{\Xint-}

\begin{document}

\title{ Theory of Eigenstate Thermalization}

\author{Tobias Helbig}
\thanks{Both authors contributed equally to this work.}
\author{Tobias Hofmann}
\thanks{Both authors contributed equally to this work.}
\author{Ronny Thomale}
\author{Martin Greiter}
\email{Corresponding author: martin.greiter@uni-wuerzburg.de}
\affiliation{Institut für Theoretische Physik und Astrophysik and Würzburg-Dresden Cluster of Excellence ct.qmat, Julius-Maximilians-Universität, 97074 Würzburg, Germany
}

\date{June 3, 2024}


\begin{abstract}
If we prepare an isolated, interacting quantum system in an eigenstate and perturb a local observable at an initial time, its expectation value will relax towards a thermal expectation value, even though the time evolution of the system is deterministic.  The eigenstate thermalization hypothesis (\ETH ) of Deutsch and Srednicki suggests that this is possible because each eigenstate of the full quantum system acts as a thermal bath to its subsystems, such that the reduced density matrices of the subsytems resemble thermal density matrices.  Here, we use the observation that the eigenvalue distribution of interacting quantum systems is a Gaussian under very general circumstances, and Dyson Brownian motion random matrix theory, to derive the \ETH\ and thereby elevate it from hypothesis to theory.  Our analysis provides a derivation of statistical mechanics which neither requires the concepts of ergodicity or typicality, nor that of entropy.  Thermodynamic equilibrium follows solely from the applicability of quantum mechanics to large systems and the absence of integrability. 
\end{abstract}
\maketitle


\section{Introduction}
\label{sec:intro}

Imagine we prepare a large, isolated quantum system in an eigenstate $\ket{\psi_n}$ of its Hamiltonian $\hat H$, and perturb it at time $t=0$ via a coupling to a local operator $\mathcal{\hat O}$.  We expect, and it has been demonstrated both in numerical~\cite{rigol_thermalization_2008,sugimoto_test_2021} and in actual experiments using cold atoms~\cite{
  ueda_quantum_2020}, that the expectation value of $\mathcal{\hat O}$ will relax towards its thermal equilibrium value, even though the time evolution of the system is governed by Schrödinger's equation, and hence deterministic.  The eigenstate thermalization hypothesis (\ETH )
of Deutsch~\cite{deutsch_quantum_1991,deutsch_eigenstate_2018} and Srednicki~\cite{srednicki_chaos_1994,srednicki_approach_1999} proposes that this is possible because each eigenstate acts as a thermal bath to its subsystems~\cite{
  rigol_thermalization_2008,
  gogolin_equilibration_2016,dalessio_quantum_2016,garrison_does_2018,cipolloni_eigenstate_2021,wang_eigenstate_2022,zhang_self-similarity_2023,adhikari_eigenstate_2023}.

More concisely, the \ETH\ states that if the entire system is in an eigenstate $\ket{\psi_n}$ and we divide it into a small subsystem S and a much larger bath B, and trace out the bath, the reduced density matrix of the system S
\begin{align}
  \label{eq:rho_n}
  \hat\rho^\S_{n} \equiv \Tr_\B \ket{\psi_n}\bra{\psi_n},
\end{align}
is equal to the thermal density matrix
\begin{align}
  \label{eq:rho_beta_n}
  \hat\rho^\S(\b) \equiv \Tr_\B \hat\rho(\b)\quad\text{with}\quad
  \hat\rho(\b)=\frac{1}{Z(\b)}\,\e^{-\b\hat H}
\end{align}
we obtain by taking the trace of a thermal density matrix for the entire system.
%
%
%
The inverse temperature $\b$ in \eqref{eq:rho_beta_n} is fixed by
\begin{align}
  \label{eq:Hb=ln}
  \braket{\hat H}_\b = -\parder*{\b}\ln Z(\b) =\l_n,
\end{align}
where $\l_n$ is the energy eigenvalue of $\ket{\psi_n}$,
$\hat H\ket{\psi_n}=\l_n\ket{\psi_n}$, and
\begin{align}
  \label{eq:Z}
  Z(\b)=\Tr \e^{-\b\hat H}
\end{align}
is the partition function of the entire system.
In other words, if the entire system is in an eigenstate of its Hamiltonian,
%
%
the local properties of this state are indistinguishable from those of a thermal state.

The \ETH\ does not hold for systems which are integrable~\cite{kinoshita_quantum_2006,rigol_relaxation_2007,li_quantum_2018}, either fully or emergent through (many body) localization~\cite{serbyn-13prl127201,huse-14prl174202,nandkishoredoi-15arcmp15,imbrie16jsp998,abanin-19rmp021001,morningstar_Avalanches_2022}, or systems with an extensive number of conservation laws~\cite{vidmar_generalized_2016}, and is not applicable to  states in proximity to closed classical orbits or integrable subsectors of otherwise not integrable models (quantum scars)~\cite{bernien_probing_2017,turner_weak_2018,serbyn_quantum_2021,Moudgalya_Quantum_2022,chandran_quantum_2023}.  Concise examples and a significant body of numerical work, however, indicate that it generally holds for interacting quantum systems whenever its validity is not restricted by conservation laws in the mentioned sense, as we assume from now on.

In this work, we derive the \ETH\ for interacting quantum systems using Dyson  Brown\-ian motion random matrix theory (\RMT)~\cite{anderson_introduction_2009,potters_first_2020}.

To begin with, we argue that the \ETH\ \eqref{eq:rho_n}--\eqref{eq:Z} is internally consistent only if the bath B is sufficiently large 
and the eigenvalue distribution of the entire quantum system, and hence the bath, is given by a Gaussian distribution
\footnote{This is not to say that closed quantum systems with other eigenvalue distributions will not thermalize, but only that \eqref{eq:rho_n}--\eqref{eq:Z}, and in particular \eqref{eq:Hb=ln}, can no longer be assumed.  As our analysis shows, the only requirement for a canonical distribution in the reduced density matrix for S is that the eigenvalue density $\rho_\B\of(E)$ of the bath in the vicinity of the relevant energy $E$ is $\propto\e^{\b E}$.  Physically, this requires that the entropy of the bath $S_\B\of(E)$ is a smooth function in this vicinity, with $\b={\partial S_\B(E)}/{\partial E}$.
}.
This is fully consistent with our assumptions since it has been shown by Hartmann, Mahler, and Hess~\cite{hartmann_spectral_2005}, and unknowingly also by ourselves using a different method~\cite{hofmann}, that for quantum systems with local interactions the eigenvalue distribution converges towards a Gaussian in the large system limit, which we assume throughout this work
  \footnote{Formally, the theorem assumes local interactions between finite dimensional Hilbert spaces at each point in the space the system is embedded in.  This yields an energy spectrum which is bounded for each unit of volume.  Our analysis, however, applies to unbounded spectra as well, since taking the limit of infinitely large Hilbert spaces at each point in space commutes with the only other limit we take, the limit of a large bath.
}.
Numerical work~\cite{hofmann} indicates that the condition of locality can be relaxed towards the requirement that the Hamiltonian consists of terms which involve far fewer degrees of freedom than the Hilbert space, like only two-, three- and four-body interactions, but to our knowledge,
this has not been shown analytically.

\begin{figure*}[t]
  \centering
  \includegraphics[width=0.97\linewidth]{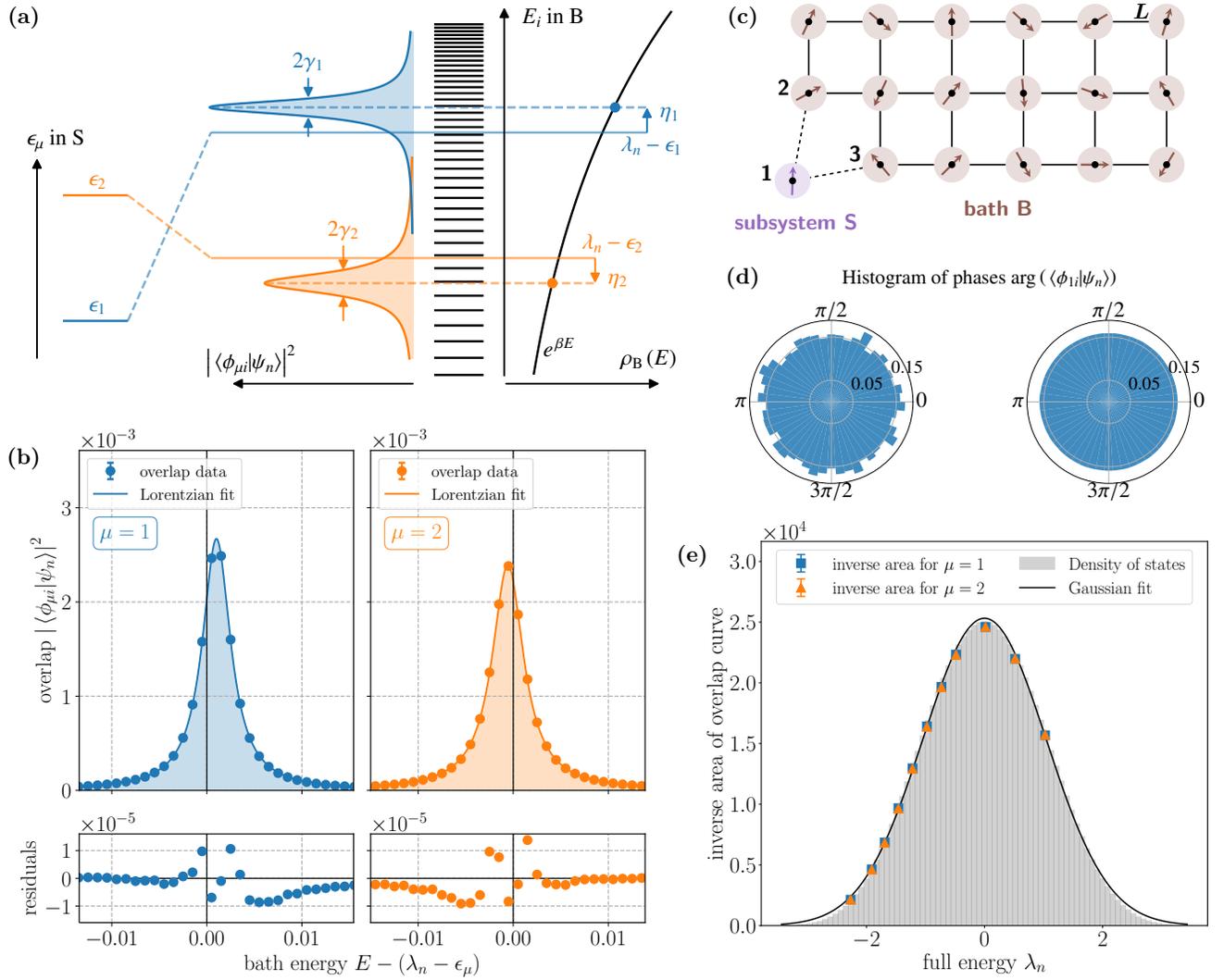}
  \caption{Analysis of overlaps $\braket{\phi_{\mu i}|\psi_n}$ of energy eigenstates $\ket{\psi_n}$ with the basis states of the unperturbed system $\ket{\phi_{\mu i}}$.
  \textbf{(a)} Illustration of the Cauchy-Lorentz distributions of the absolute squared overlaps in terms of the bath energies $E_i$ with widths $\gamma_\mu$ and shifts $\eta_\mu$. 
  \textbf{(b)} Numerical data of the absolute squares of the overlaps $\abs{\braket{\phi_{\mu i}|\psi_n}}^2$ for $\mu=1$ in blue and $\mu=2$ in orange, in a system with one spin in the subsystem S and 17 spins in the bath B. 
  The peaks can be described rather accurately by Cauchy-Lorentz distributions, as indicated by the solid curve fit and the small fit residuals shown below.
  \textbf{(c)} The spin $\half$ lattice for S and B, here with 18 sites.  S consists only of site 1, and the region R of sites 2 and 3.
  \textbf{(d)} Normalized histograms of the phases of the overlaps $\arg\left(\braket{\phi_{1 i}|\psi_n}\right)$ for $\mu=1$, computed in terms of 50 intervals from the data of (b).
  Left: single eigenstate. Right: average over $N_\text{av} = 3200$ states $\ket{\psi_n}$. 
  \textbf{(e)} Plot of the inverse of the integrated area $A_{\mu ,n}$ of the overlap curves (see \eqref{eq:def_A_mu_n_via_Lorenzian}) for different energies $\l_n$ in a system with 15 sites in B and one site in S. 
  We find $A_{1,n}=A_{2,n}$ for each $n$.  The inverse areas $1/A_{1,n}=1/A_{2,n}$ are equal to the numerically obtained eigenvalue densities $\rho_\B({\l_n})$ of the bath, which we approximate as a Gaussian curve with standard deviation $\D_\B = \num{1.0438 \pm 0.0033}$.}
  \label{fig1}
\end{figure*}

In most of the analysis thereafter, we consider random systems where the coupling $\hat X$ of the small subsystem S and the bath B is weak, and can hence be treated perturbatively.  We further assume that the bath is large, and that $\hX$ couples only to a small region R of the bath.  In this setting, we study the decomposition of exact eigenstates of the entire system $\ket{\psi_n}$ in terms of basis states of the unperturbed system,
\begin{align}
  \label{eq:3}
  \ket{\phi_{\mu i}}\equiv \ket{\phi_{\mu}^\S}\otimes \ket{\phi_i^\B},
\end{align}
\ie direct products of eigenstates of the subsystem
$\hat H_\S\ket{\phi_\mu^S}=\eps_\mu\ket{\phi_\mu^S}$, $\mu=1,\ldots ,N_\S$, and the bath
$\hat H_\B\ket{\phi_i^B}=E_i\ket{\phi_i^B}$, $i=1,\ldots ,N_\B$.  Our numerical work, and our analytic calculations using 
\RMT , show that the statistical expectation values
of the squares of the overlaps
$\chi_{\mu i,n}\equiv\E\bigof[\abs{\!\braket{\phi_{\mu i}|\psi_n}\!}^2]$ are given by Cauchy-Lorentz distributions
\begin{align}
  \label{eq:4}
  \chi_{\mu i,n} 
  \approx\frac{1}{\pi}
  \frac{\g_\mu}{\of(\eps_\mu+E_i-\l_n-\eta_\mu)^2+\g_\mu^2},
\end{align}
(see Fig.~\ref{fig1}\figletter{a}).  The peaks of the distributions, when plotted as functions of the energies $E_i$ of the basis states of the bath, have width $\g_\mu$ and are shifted by $\eta_\mu$ relative to the positions $\l_n-\eps_\mu$ they assume when the coupling is infinitesimal.

To first order in the matrix variance (or second moment) $t$ of the perturbation $\hat X$,
\begin{align}
  \label{eq:mat_var_X}
  t=\frac{1}{N}\,\E\of[\,\Tr\of(\hat X^2)\,],
\end{align}
where 
$N=N_\S\hspace{0pt} N_\B$ is the matrix  dimension of $\hat X$, we find:
%
%
\begin{enumerate}

\item The half-widths of the Lorentzians are given by
  \begin{align}
    \label{eq:intro_gamma_mu}
    \phantom{oooo}\g_\mu=t\frac{\spi}{\D_0}
    \sum_\nu \s_{\mu\nu}^2\e^{-\teps_{\mu\nu}^{\,2}},
  \end{align}
  where $\teps_{\mu\nu}\equiv\frac{1}{\D_0}\Of{\eps_{\mu}-\eps_{\nu}}-b_0$, $b_0 \equiv\frac{1}{4}\b\D_0$, and $\D_0$ is a parameter which roughly corresponds to the bandwidth of the region R of the bath the interaction matrix $\hX$ couples to (see \eqref{eq:tau_ij} and \appendixref{app:tau_variances} below).  The parameters $\s_{\mu\nu}^2$ reflect that the perturbation $\hX$ does, in general, not indiscriminately scatter between all the different levels $\eps_\mu$ and $\eps_\nu$ of the system.
%

\item The shifts are given by
  \begin{align}
    \label{eq:intro_eta_mu}
    \phantom{oooo}\eta_\mu
    &=-t\frac{\spi}{\D_0} \sum_\nu \s_{\mu\nu}^2
      \e^{-\teps_{\mu\nu}^{\,2}} \,\erfi\of(\teps_{\mu\nu})
    \nonumber\\[3pt]
    &=-t\frac{2}{\D_0} \sum_\nu \s_{\mu\nu}^2
      \biggl(\teps_{\mu\nu}-\frac{2\teps_{\mu\nu}^{\;3}}{3}
      +\frac{4\teps_{\mu\nu}^{\;5}}{15}-\ldots\biggr),
  \end{align}
  where $\erfi(z)\equiv -\i\,\erf(\i z)$ is the imaginary error function 
({Erfi}$[\mathinner{x}]$ in \emph{Mathematica}).
They reflect an effective level repulsion, due to the coupling $\hX$.

\item \label{it:hatrho} While the widths and heights of the Lorentzians depend on the index $\mu$, the integrated areas beneath them do not.  Since expansion of the Gaussian eigenvalue density of the bath around any specific value for the energy yields an exponential, $\rho_\B(E)\propto \e^{\b E}$, the reduced density matrix of the system S has diagonal entries
\begin{align}
  \label{eq:10}
  \phantom{oooo}\bra{\phi_{\mu}^\S} \hat\rho_n^\S \ket{\phi_{\mu}^\S}
  =\frac{1}{Z^\S} \e^{-\b\,\of(\eps_{\mu}-\eta_\mu)},
\end{align}
where $Z^\S=\sum_\mu \e^{-\b\,\of(\eps_{\mu}-\eta_\mu)}$ is the partition function of the small subsystem in the presence of the coupling $\hX$.  The off-diagonal entries vanish as $\sfrac{1}{\sqrt{N_\B}}$,
and hence exponentially, with the size of the bath B.

\item 

The reduced density matrix of S is hence that of a canonical distribution with (inverse) temperature $\b$ and (for weak coupling infinitesimally) shifted energy levels $\eps_{\mu}-\eta_\mu$.  When we  spectroscopically observe level spacings in S, we likewise do not observe the bare levels $\eps_{\mu}$, but the shifted levels $\eps_{\mu}-\eta_\mu$.  Therefore, it is only appropriate that they, and not the bare levels, enter the distribution.  For a transition from level $\mu$ to $\nu$, the lineshape is a 
Lorentzian with half-width $\g_\mu+\g_\nu$.


\end{enumerate}


%
Finally, we will show in Sec.~\ref{sec:generalization} with \appendixref{app:validity}, and the consistency condition studied in Sec.~\ref{sec:consistency} and \appendixref{app:nesting}, that our weak coupling analysis for random interactions between S and B implies the validity of the \ETH\ in general.  This includes situations where the coupling between S and B is neither weak nor random.
As a corollary, our analysis provides a derivation of statistical mechanics which requires neither
ergodicity~\cite{moore_ergodic_2015} nor typicality~\cite{tasaki_quantum_1998,goldstein_canonical_2006, popescu_entanglement_2006,gogolin_equilibration_2016}, and hence supersedes the historical derivations~\cite{Kubo65}.


For sufficiently large quantum systems, the assumption of randomness in our calculation is redundant due to self averaging or concentration of measure.  The theorem here is that in the limit of large matrix dimensions%
, $\E\of[\,\Tr\of(f(\hat M))\,]=\Tr\of(f(\hat M))$, where $\E\of[\, .\,]$ is the statistical expectation value, $f$ a sufficiently smooth function, and $\hat M$ a random matrix with statistically independent entries~\cite{anderson_introduction_2009}.  In our context, the method of \RMT\ does hence not entail a statistical assumption, but merely provides a technique to extract analytic information about large systems.


\section{Consistency requirements}
\label{sec:consistency}

Let us assume that the bath B, and hence the entire system, is very large, and recall that the (normalized) eigenvalue density is given by a Gaussian~\cite{hartmann_spectral_2005,hofmann},
\begin{align}
  \label{eq:rho_l}
  \rho\of(\l )=\sqrt{\frac{\a }{2\pi }}\,\e^{-\half\a \l^2},
\end{align}
where 
$\frac{1}{\a } =\D^2_{\tot}$
is the matrix variance of the Hamiltonian $\hH$ of the entire system. 
Expansion around $\l_n$ yields
\begin{align}
  \label{eq:bn=-a_ln}
  \rho\of(\l )\propto \e^{\b_n\l} \quad\text{with}\quad
  \b_n=\parder{\ln \rho\of(\l )}{\l}_{\l=\l_n}=-\a\l_n.
\end{align}
According to \ref{it:hatrho} above, this temperature enters the canonical distribution for the subsystem S.  It is positive for $\l_n<0$, which we assume in the following.  (Our analysis, however, applies equally to $\l_n>0$, where $\b_n<0$.)

If the \ETH\ is to hold, the temperature $\b_n$ will also be connected to $\l_n$ via \eqref{eq:Hb=ln}.  With
\begin{align}
  \label{eq:Zb}
  Z(\b)=N\intd{\diff\l} \rho\of(\l ) \e^{-\b\l} = N \exp\Of{\frac{\b^2}{2\a}}
\end{align}
we find
\begin{align}
  \label{eq:ln=-bn:a}
  \l_\n=\braket{\hat H}_{\b_n}
  = -\parder*{\b}{\ln Z (\b)}\biggl|_{\b=\b_n}\biggr. = -\frac{\b_n}{\a},
\end{align}
which is equivalent to \eqref{eq:bn=-a_ln}.  The \ETH, as formulated in \eqref{eq:rho_n}--\eqref{eq:Z} above, is hence consistent if the eigenvalue distribution is a Gaussian, which in turn is always the case for locally interacting quantum systems.  Note that this consistency requires a relation between a weighted integral over $\rho\of(\l )$ and its (logarithmic) derivative at $\l=\l_n$.
To our understanding, this condition specifies the functional form of $\rho\of(\l )$ up to the parameter $\a$ and an overall normalization.

An even more stringent consistency condition can be derived from nesting.
Since the thermal state $\hat\rho\of(\b)$ in \eqref{eq:rho_beta_n} for the entire system consists of a sum of eigenstates $\ket{\psi_n}\!\bra{\psi_n}$, we can apply the \ETH\ to each of them.  This yields a sum of canonical distributions $\hat\rho^\S\of(\b_n)$ for 
the subsystem S, which are weighted with the Boltzmann factor $\e^{-\b\l_n}$ for the entire system.  Consistency requires now that this weighted sum of exponential functions with different temperatures $\b_n$ must be equal to a single exponential $\hat\rho^\S\of(\b)$.  In \appendixref{app:nesting}, we show that this is indeed the case if we assume the Gaussian eigenvalues density \eqref{eq:rho_l} and take the limit of a large bath.

\section{Numerical results}
\label{sec:numerics}


We now describe our numerical work, which has initially inspired, and was subsequently guided by, the analytical solution reported in Sect.~\ref{sec:analytics} below.  Many of the results presented here can only be fully appreciated in the context of the analytical solution.

As mentioned in the introduction, we consider a quantum system consisting of a small subsystem S and a much larger bath B, which are weakly coupled through a perturbation $\hX$.  %
%
We consider exact eigenstates $\ket{\psi_n}$ of the total Hamiltonian $\hH=\hH_\S+\hH_\B+\hX$, and investigate the overlaps of those with the eigenstates $\ket{\phi_{\mu i}}$ of the unperturbed system.
As a numerically accessible example of such a system with local interactions and a high amount of randomness, we study a spin-$\half$ lattice with up to 18 sites,
where we (mostly) represent the subsystem S by a single spin $\hs_1$ on site 1,
as illustrated in Fig.~\ref{fig1}\figletter{c}.  The Hamiltonian for a system with $L$ sites is given by
\begin{align}
  \label{eq:num_ham}
  \hH=h_1\hs_1^3+\sum_{\langle ij\rangle}\sum_{\a,\b=1}^3J_{ij}^{\a\b}\hs_i^\a \hs_j^\b
  +\sum_{\a=1}^3 h_L^\a\hs_L^\a,
\end{align}
where $\hs_i^\a$ with $\a=1,2,3$ are the three Pauli matrices. 
The first term in
\eqref{eq:num_ham}
represents $\hH_\S$, and splits the two levels by a fixed onsite magnetic field $h_1=\half\of{\eps_2-\eps_1}$.  The second term contains a sum over all the links $\langle ij\rangle$ between neighboring sites, as indicated in Fig.~\ref{fig1}\figletter{c}.
To maximize the randomness and allow for as little symmetry as possible, we sum over all nine possible combinations of Pauli matrices with independent coefficients $J_{ij}^{\a\b}$ drawn from Gaussian distributions $\N(0,\k_{ij}^2)$ (see \eqref{eq:p_of_x} below for the definition).  To break the $\mathbb{Z}_2$ spin reflection symmetry of the second term, we add the third.  It describes an onsite magnetic field $\vec h_L$ in a random direction on site $L$ (site 18 for an 18 site lattice) far away from the subsystem S. The three coefficients $h_L^\a$ are likewise drawn from Gaussian distributions $\N(0,\k_{L}^2)$.

The second term in \eqref{eq:num_ham} contains both the interaction $\hX$ between S and B (terms with $i=1$, $j\in\RR$) and the interactions within B (terms with $1<i<j$).  In our numerics, we choose the variances $\k_{ij}^2$ and $\k_{L}^2$ of all the terms in $\hH_\B$ equal and normalized such that the total matrix variance is
\begin{align}
  \label{eq:mat_var_B}
  \D_\B^2\equiv\frac{1}{N}\,\E\bigl[\,\Tr\of{\hH_\B^2}\bigr]=1,
\end{align}
where $\hH_\B$ is written as a matrix with 
dimension $N$.  We further choose the variances $\k_{1j}^2$ of all contributions to $\hX$ equal and normalized such that the matrix variance \eqref{eq:mat_var_X} of $\hX$ is $t$.  For the matrix variance of $\hH_\S$, we choose $\D_\S^2=h_1^2=0.01$.

To diagonalize \eqref{eq:num_ham} for specific samples of random couplings we use the algorithm of multiple relatively robust representations (MRRR)~~\cite{dhillon_multiple_2004} within the LAPACK package~~\cite{anderson_lapack_1999} as a direct eigensolver to compute the entire spectrum in systems with up to 15 sites.
For larger system sizes, we use the FEAST algorithm~~\cite{polizzi_density-matrix-based_2009,peter_tang_feast_2014}, which maps the eigenvalue problem to a complex contour integral of the Green's function and is able to determine the eigensystem in any subspace without knowledge of the remainder of the spectrum.

In Fig.~\ref{fig1}\figletter{a}, we schematically illustrate the overlap curves $\abs{\braket{\phi_{\mu i}|\psi_n}}^2$ plotted as functions of the bath energies $E$ for a two-state subsystem S as indicated by our numerical results.  The overlaps display sharp peaks with half-width $\g_{\mu,n}$ for each level $\eps_\mu$ of S centered around $E=\l_n-\eps_\mu+\eta_{\mu,n}$.  
To obtain meaningfully smooth curves, we have to average over thousands of random samples.  Due to the computational complexity, it proved intractable to diagonalize a sufficiently large number of Hamiltonians with random interactions.
Since eigenstates with energies nearby yield very similar results for the overlap curves, we could circumvent this problem by averaging $\abs{\braket{\phi_{\mu i}|\psi_m}}^2$ over 3200 states $\ket{\psi_m}$ in a narrow interval around a chosen energy $\l_n$.  The results for an 18 site system with $t=\num{6.25e-4}$ and $\l_n = \num{-0.95789 \pm 0.00025}$ are shown in Fig.~\ref{fig1}\figletter{b}.  The overlap curves close to their peak are rather accurately described by Cauchy-Lorentz distributions, as illustrated by the nonlinear regression fits on the data using the method of nonlinear least squares shown below the peaks. Further data of the overlaps is shown in \appendixref{app:lorentzian}.
From the Lorentzian fits, we extract $\g_{\mu,n}$ and $\eta_{\mu,n}$, yielding $\gamma_{1,n} = \num{1.804e-3}$, $\gamma_{2,n} = \num{2.013e-3}$, $\eta_{1,n} = \num{1.013e-3}$ and $\eta_{2,n} = \num{-0.565e-3}$.
As one can see from Fig.~\ref{fig1}\figletter{b}, the widths and shifts of the Lorentzians depend on $\mu$.
To account for the statistical variation in the fit parameters due to the random Hamiltonians, and to trace their functional dependence on $t$ and $\l_n$, we take an additional average over hundreds of samples of $\hX$ and $\hH_\B$.  The results are shown and discussed in Fig.~\ref{fig3}\figletter{a} and \figletter{b}.

In Fig.~\ref{fig1}\figletter{d}, we plot the phases for the overlaps $\braket{\phi_{1i}|\psi_n}$ for the data of the blue $\mu=1$ peak of Fig.~\ref{fig1}\figletter{b}, on the left for different basis states $\ket{\phi_{\mu i}}$ and a single eigenstate $\ket{\psi_n}$, and on the right as an average over the 3200 eigenstates $\ket{\psi_m}$. The resulting probability distribution is uniform, and thereby validates the assumption of random phases implicit in \RMT.


\begin{figure*}[!t]
  \centering
    \includegraphics[width=.9\linewidth]{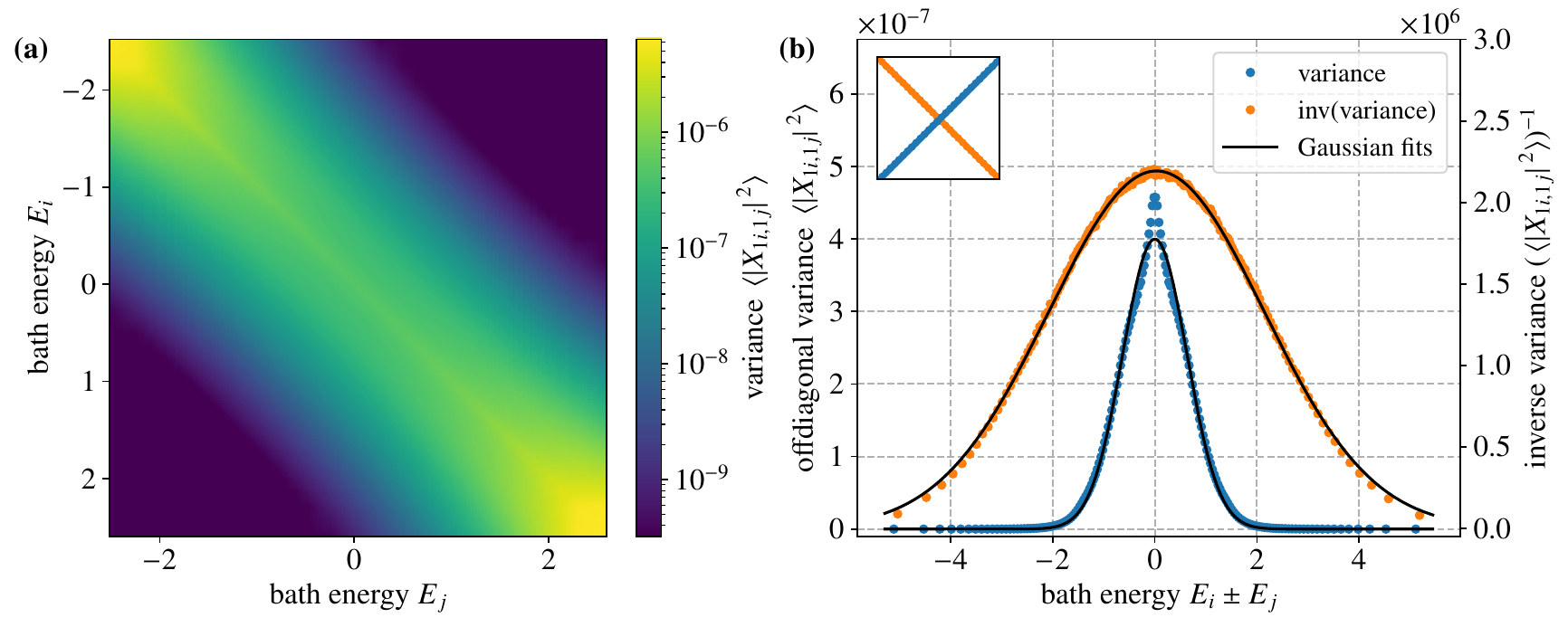}
  \caption{Analysis of the coupling matrix $\hX$ between a subsystem S with 1 site and a bath B with 12 sites in the unperturbed basis $\ket{\phi_{\mu i}}$ of S and B.
  \textbf{(a)} Heat map representation with Gouraud shading type of the off-diagonal variance $\sigma_{1 i,1 j}$ of $\hX$ for $(\mu,\nu) = (1,1)$ plotted as a function of the mean bath energies $E_i$ and $E_j$, averaged over boxes with $(25 \times 25)$ elements.  The variance along the diagonal direction of the matrix increases towards the edges, and falls off in the perpendicular direction.  The band character of the matrix $\hX$ with a well-defined, energy-independent scattering width $\D$ is visible.
  \textbf{(b)} Plot of the variance along the anti-diagonal direction of $\hX$ as a function of $E_i-E_j$ in blue and the inverse of the off-diagonal variance along the diagonal direction as a function of $E_i+E_j$ in orange. The data are normalized such that the maximum is equal to one for each set of data.  Gaussian fits for both directions are shown in black, yielding standard deviations of $\D_{\text{d}} = \num{2.1099 \pm 0.0035}$ and $\D_{\text{ad}} = \num{0.6259 \pm 0.0045}$ for the orange and blue data, respectively.  The data points in the off-diagonal direction show an increased peak around zero, which we attribute to finite size effects.}
  \label{fig2}
\end{figure*}


To obtain Fig.~\ref{fig1}\figletter{e}, we fit the numerical data for the overlaps $\abs{\braket{\phi_{\mu i}|\psi_n}}^2$ for a 16 site lattice with $t=\num{2.5e-3}$ to
Lorentzians
\begin{align}
  \label{eq:def_A_mu_n_via_Lorenzian}
  f_{\mu,n}\of{E_i}= A_{\mu,n}\frac{1}{\pi}  \frac{\g_{\mu,n}}
  {\left(E_i+\eps_\mu-\l_n-\eta_{\mu,n}\right)^2+\left(\g_{\mu,n}\right)^2},
\end{align}
where $\g_{\mu,n}$, $\eta_{\mu,n}$, and the integrated areas $A_{\mu,n}$, are fit parameters.  We average these over 400 eigenstates $\ket{\psi_n}$ with energies close to $\l_n$, 100 samples of $\hH_\B$, and  two samples of $\hX$ for each $\hH_\B$. In Fig.~\ref{fig1}\figletter{e}, we plot the reciprocal values of 
$A_{\mu,n}$ for various energies $\l_n$.  We find that, first, the areas are independent of $\mu$, and second, that they are indirectly proportional to the (Gaussian) eigenvalue density $\rho(\l_n)$ of the entire system.
In the notation of Sect.~\ref{sec:analytics}, this result is equivalent to the condition \eqref{eq:sum_n_chi}.
It thereby confirms the validity of Ansatz \eqref{eq:chi_mu_i_n_ansatz}, which, as we will show in Sect.~\ref{sec:analytics}, implies a canonical distribution for the reduced density matrix $\hat\rho^\S_{n}$ of the subsystem S.


In Fig.~\ref{fig2}, we investigate the structure of the matrix $\hX$ with $t=\num{2.5e-3}$ in the full, unperturbed basis $\ket{\phi_{\mu i}}$.
Since $\hX$ scatters spins in a small region R of B, it only scatters between basis states $\ket{\phi_{i}^\B}$ which are sufficiently close in energy, independently of whether it flips the spin $\hs_1$ in S or not.
We hence expect the variances $\s_{\mu i, \nu j}^2$ of the elements to factorize according to \eqref{eq:sigma}, $\s_{\mu i,\nu j}^2 =\s_{\mu\nu }^2\hspace{1pt} \t_{ij}^2$.
In our numerics, S consists of only two levels, and since there are twice as many off-diagonal terms (those with $\hs^1,\hs^2$ acting on site 1) than diagonal terms (those with $\hs^3$ acting on site 1) in our interaction, we expect $\s_{11}^2=\s_{22}^2=\frac{1}{3}$, $\s_{12}^2=\s_{21}^2=\frac{2}{3}$.

In Fig.~\ref{fig2}\figletter{a}, we show 
$\s_{1i,1j}^2$ for an average over $100$ random Hamiltonians $\hH_B$ and two random interactions $\hX$ for each $\hH_B$ as a heat map plotted over the bath energies $E_i$ and $E_j$. 
The heat maps for 
$(\mu,\nu)=(1,2),\,(2,1)$ and $(2,2)$ shown in \appendixref{app:xmatrix} look identical
and confirm the factorization with the above values for $\s_{\mu\nu }^2$.
%
In Fig.~\ref{fig2}\figletter{b}, we show that $\t_{ij}^2$ is indirectly proportional to the eigenvalue density $\rho_0(E)$ of the bath along the diagonal $E_i=E_j$, and can be approximated by a Gaussian in $E_i-E_j$ when plotted along the anti-diagonal $E_i+E_j=0$.
This is illustrated by the Gaussian fits in Fig.~\ref{fig2}\figletter{b}, which yield standard deviations of $\D_{\text{d}} = \num{2.1099 \pm 0.0035}$ and $\D_{\text{ad}} = \num{0.6259 \pm 0.0045}$ for the diagonal and anti-diagonal direction, respectively. $\D_{\text{d}}/2$ agrees with the standard deviation of $\rho_0(E)$.
We attribute the discrepancy of the peak along the anti-diagonal at $E_i-E_j=0$ to the Gaussian fit to finite-size effects, \ie the fact that the region R consists of only two lattice sites (see Fig.~\ref{fig1}\figletter{c}).
With $\D_0=\sqrt{2}\,\D_{\text{ad}}=\num{0.8852 \pm 0.0064}$, the variances of the matrix $\hX$ are consistent with the analytical result \eqref{eq:tau_ij}. 
In \appendixref{app:xmatrix}, we further elaborate our numerical results and their agreement with the analytical result.

\begin{figure*}[!t]
  \centering
  \includegraphics[width=0.97\linewidth]{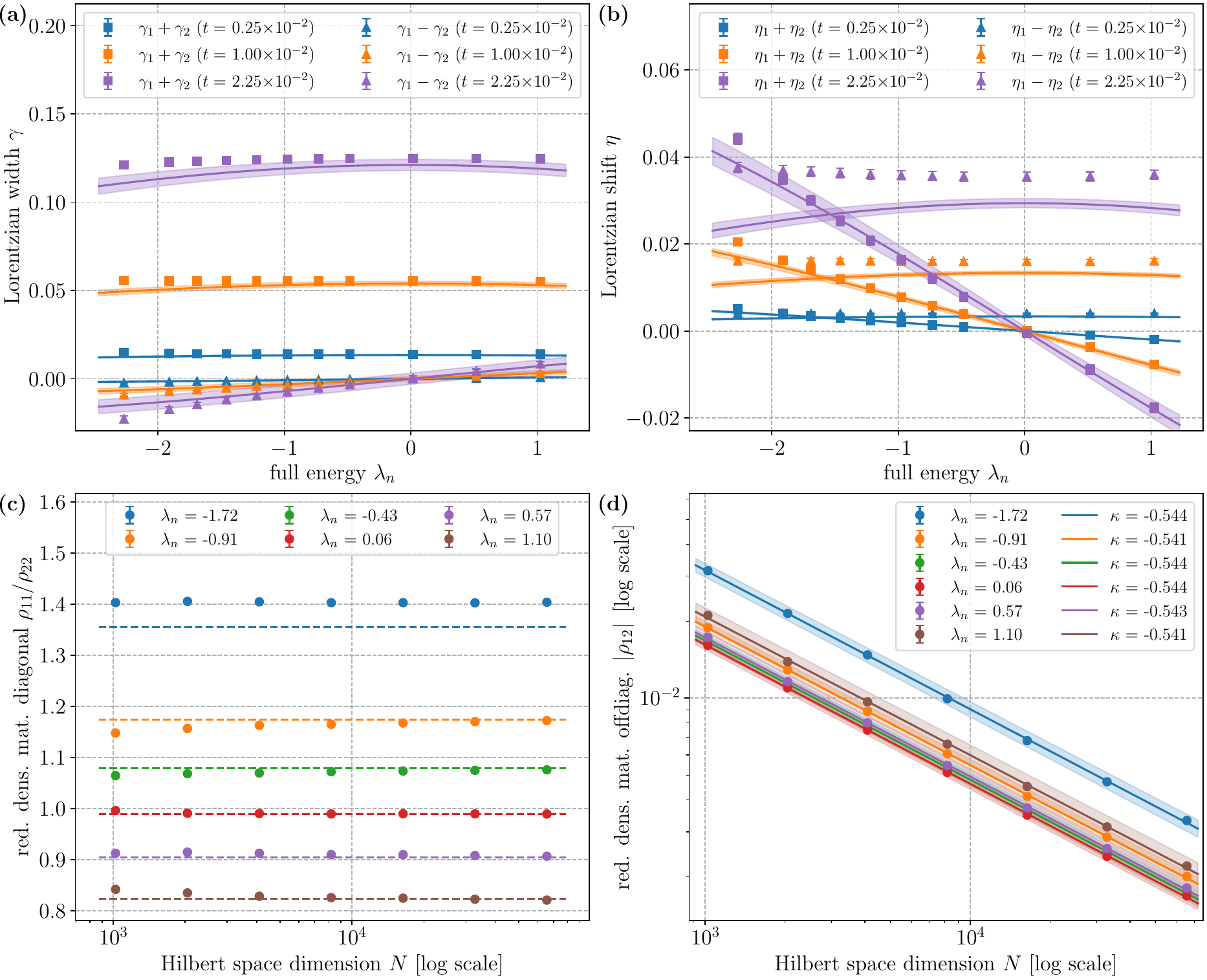}
  \caption{Analysis of the Lorentzian parameters and the reduced density matrix.
    \textbf{(a)} Lorentzian half-width $\g_{\mu,n}$ (\cf \eqref{eq:intro_gamma_mu}) for the two peaks of a system with 1 site in S and 15 sites in B, plotted as the sum $\g_1+\g_2$ and the difference  $\g_1-\g_2$ for various energies $\l_n$.
    \textbf{(b)} Shifts $\eta_{\mu,n}$ (\cf \eqref{eq:intro_eta_mu}) for the same system, again plotted as $\eta_1+\eta_2$ and $\eta_1-\eta_2$.  In both plots, the analytical results 
    are shown as solid lines, with error intervals originating from the fit error of the parameters inserted into \eqref{eq:intro_gamma_mu} and \eqref{eq:intro_eta_mu}.
    \textbf{(c)} Ratio of the diagonal values $\rho_{11}$ and $\rho_{22}$ of the reduced density matrix in a two-state subsystem S  for different energies $\l_n$, plotted as a function of the total Hilbert space dimension $N$.
  The dashed line represents the theoretical expectation obtained from the inverse temperature $\beta$, as computed from the eigenvalue density (see \eqref{eq:bn=-a_ln}).
  \textbf{(d)} Absolute value of the off-diagonal element $\abs{\rho_{12}}$ of the same reduced density matrix as in \textbf{(c)}, plotted on a log-log scale. Linear fits to the data points show a power law decline with an exponent around to $\k=-0.5$, as predicted by \eqref{eq:rho_mu_nu_squared}.
  }
  \label{fig3}
\end{figure*}

In Figs.~\ref{fig3}\figletter{a} and \figletter{b}, we show the parameters
$\g_{\mu,n}$ and $\eta_{\mu,n}$ obtained through fitting for various energies $\l_n$ and interaction strengths $t$, or more precisely, the sums and differences of these values for the two peaks, $\g_1\pm \g_2$ and $\eta_1\pm \eta_2$, for each value of $\l_n$.  The data are obtained from a 16 site lattice by averaging over 400 eigenstates $\ket{\psi_n}$ in an energy interval close to $\lambda_n$, 100 samples of $\hH_B$ and two samples of $\hX$ for each $\hH_B$.
The analytical values for $\g_1\pm \g_2$ and $\eta_1\pm \eta_2$, as calculated in Sect.~\ref{sec:analytics} and given in~\eqref{eq:intro_gamma_mu} and~\eqref{eq:intro_eta_mu}, are indicated by solid curves
and are comparable to our the numerical results.  We attribute the deviations for large negative $\l_n$ to finite size effects amplified by the low eigenvalue density for large values of $\abs{\l_n}$.  The analytical result depends on the scattering bandwidths of $\hX$.  The value $\D_0= \num{0.6131 \pm 0.0024}$ was obtained by a Gaussian fit to the variances of the elements of $\hX$, as shown in Fig.~\ref{fig2}\figletter{b}.  The standard error of $\D_0$ results in a margin of error for our analytical result, which we indicate through the opaque regions in Figs.~\ref{fig3}\figletter{a} and \figletter{b}.  The data confirm that both
$\g_{\mu}$ and $\eta_{\mu}$
scale in first order linearly with $t$, as detailed in \appendixref{app:lorentzian}.

In Fig.~\ref{fig3}\figletter{a}, we see that the sum of the peak widths $\g_1+\g_2$ depends only weakly on $\l_n$, while the difference $\g_1-\g_2$ is roughly proportional to $\l_n$.  The finite width of the peaks can be attributed to the scattering between the two levels $\eps_1$ and $\eps_2$ induced by $\hX$.  Since the eigenvalue density of the bath decreases towards the spectral boundaries, the peak closer to the boundary shows a larger broadening.
In Fig.~\ref{fig3}\figletter{b}, we see that the situation is reversed for the shifts.  Here, the sum $\eta_1+\eta_2$ is roughly proportional to $\l_n$.  It represents an overall shift of both peaks towards the spectral boundary, and reflects that the perturbation $\hX$ increases the overall matrix variance of $\hH$, and thus the width of the Gaussian eigenvalue distribution.  Since the inverse temperature $\b$ is proportional to the energy $\l_n$ (see \eqref{eq:bn=-a_ln}), we refer to this contribution as the thermal shift.  The difference $\eta_1-\eta_2$ depends only weakly on $\l_n$.  It reflects an effective level repulsion induced by $\hX$.  We refer to it as the repulsion shift.
The relative size of thermal and repulsion shifts depends on $t$ and $\b$, and one cannot say in general which contribution is dominant.  All in all, there is reasonable agreement between our numerical experiments for very small system sizes and our analytical calculations, where we assume small couplings $t$ and an infinitely large bath B.

In Figs.~\ref{fig3}\figletter{c} and \figletter{d}, we numerically evaluate the canonical distributions as given by the reduced density matrices $\hat\rho^\S_{n}$.  In Fig.~\ref{fig3}\figletter{c}, we plot the ratio of the diagonal entities $\rho_{11}/\rho_{22}$ of $\hat\rho^\S_{n}$ 
in systems with $L=\{ 10, 11, 12, 13, 14, 15, 16 \}$ lattice sites and corresponding interaction strengths $t=\{\num{5.92}, \num{5.08}, \num{4.44}, \num{4.10}, \num{3.77}, \num{3.36}, \num{3.13}\} \times 10^{-3}$,
where we average over $N_{\text{av}} =\{ 50, 100, 200, 400, 800, 1600, 3200 \}$ states $\ket{\psi_n}$ in a narrow energy window around the respective values of $\l_n$.  We additionally average over 100 samples of $\hH_B$ and two samples of $\hX$ for each $\hH_B$.
We see that the ratios $\rho_{11}/\rho_{22}$ are independent of the Hilbert space dimension $N$ of the entire system,
and that they roughly match the theoretical predictions for Boltzmann factors with $\b= -\l_n/\D_{\tot}^2$, $\D_{\tot}^2=\D_\B^2+t+h_1^2$ (\cf \eqref{eq:bn=-a_ln}), where $\D_\B^2$ with $\D_\B=\num{1.044}$ is the variance of the normalized eigenvalue density $\rho_\B$ of the bath taken from the numerically observed distribution.
In Fig.~\ref{fig3}\figletter{d}, we plot the absolute values of the off-diagonal entities $\abs{\rho_{12}}$ using a log-log scale.  The slopes of the lines we fit for each value of $\l_n$ average to $\bar{\k} = \num{-0.543}$, in reasonable agreement with our analytical result $\k = -0.5$ in \eqref{eq:rho_mu_nu_squared}.  The numerical values for the slopes for different $\l_n$ barely deviate from $\bar{\k}$, which suggests a universal scaling exponent.  At present, we do not understand the origin of the deviation from our calculation.
The variances of the entries in the reduced density matrix scale as $1/N$ with the Hilbert space dimension for both diagonal and off-diagonal entities, as detailed in \appendixref{app:red_dens_mat}.

%

\section{Eigenvector percolation in RMT }


We begin with a summary of Dyson Brownian motion
\RMT~\cite{potters_first_2020},
and derive a convenient formula for the 
overlaps of eigenvectors as the eigenvector percolation proceeds.

The general problem is the perturbation of a diagonal $N\times N$ matrix $A$ by a Hermitian band matrix $X^t$ with random Gaussian entities,
\begin{align}
  \label{eq:11}
  M^t = A + X^t
\end{align}
where $A\ket{\phi_{\mu i}}=a_{\mu i}\ket{\phi_{\mu i}}$.  (For the review of the formalism, we could have replaced $\mu,i$ by a single index.)  The off-diagonal entities of $X^t$ are given by
\begin{align}
  \label{eq:9}
  &X_{\mu i,\nu j}^t=\frac{\s_{\mu i,\nu j}}{\sqrt{2}}
  \Of{B_{\mu i,\nu j}'(t)+\i B_{\mu i,\nu j}''(t)},\\
  &X_{\nu j,\mu i}^t= \left(X_{\mu i,\nu j}^t\right)^*
\end{align}
for $\mu<\nu$ or $\mu=\nu\ \wedge\ i<j$, and the diagonal entities by
\begin{align}
  \label{eq:8}
    X_{\mu i,\mu i}^t=\s_{\mu i,\mu i} B_{\mu i}(t),
\end{align}
where all the $B',B'',B$'s are independent Brownian motions, \ie random variables with Gaussian probability distributions $\N(0,t)$, 
\begin{align}
  \label{eq:p_of_x}
  \P[x]=\frac{1}{\sqrt{2\pi t}}\exp\Of{-\frac{x^2}{2t}},
\end{align}
and hence with variance $\E[x^2]=\intd{\diff x} \P[x]\, x^2=t$.
The resolvent
\begin{align}
  \label{eq:12}
  R^t(z)\equiv \frac{1}{z-M^t}
\end{align}
is a matrix-valued function with poles in the complex variable $z$ on the real axis at the eigenvalues of $M^t$.  With the eigenbasis 
for the perturbed system,
$M^t\ket{\psi_n}=\l_n\ket{\psi_n}$, we write the diagonal elements
\begin{align}
  \label{eq:R_diagonal}
  R_{\mu i}^t(z)\equiv \braket{\phi_{\mu i}|R^t(z)|\phi_{\mu i}}
  =\sum_n \frac{\bigabs{\braket{\phi_{\mu i}|\psi_n}}^2}{z-\l_n}
\end{align}
and obtain the (sample) Stieltjes transformation
\begin{align}
  \label{eq:15}
  g^t(z)=\frac{1}{N}\Tr R^t(z) =\frac{1}{N}\sum_n \frac{1}{z-\l_n},
\end{align}
which in the limit of infinite matrices converges to a deterministic function,
\begin{align}
  \label{eq:16}
  \G^t(z)=\lim_{N\to\infty} g^t(z).
\end{align}
For the Gaussian unitary ensemble,
$\s_{\mu i,\nu j} =\frac{1}{\sqrt{N}}$,
$\G^t(z)$ determines its own time evolution through Burgers equation,
\begin{align}
  \label{eq:Burgers}
  \bigof{ \partial_t + \G^t(z) \partial_z} \G^t(z)&=0
 \\  \rightarrow \ \G^t(z)&=\G^0(z-t\G^t(z)),
\end{align}
or more generally the time evolution of the (statistical expectation value of the) resolvent $\R^t(z)\equiv\E\of[R^t(z)]$ through the Bouchaud--Potters equation~\cite{bouchaud_two_2018},
\begin{align}
  \label{eq:BouchaudPotters}
  \bigof{ \partial_t + \G^t(z) \partial_z} \R^t(z)&=0
  \\\rightarrow\ \R^t(z)&=\R^0\bigof{z-t\G^t(z)}.
\end{align}
At $t=0$, $M^t=A$, and the diagonal elements of the resolvent are given by
\begin{align}
  \label{eq:18}
  \R_{\mu i}^0(z)=
  \E\Of[\braket{\phi_{\mu i}|\frac{1}{z-A}|\phi_{\mu i}}]=\frac{1}{z-a_{\mu i}},
\end{align}
such that \eqref{eq:BouchaudPotters} implies
\begin{align}
  \label{eq:CasatiGirko}
  \R_{\mu i}^t(z)=\frac{1}{z-a_{\mu i}-t\G^t(z)}.
\end{align}
Casati and Girko~\cite{casati_wigners_1993}, and Shlyakhtenko~\cite{shlyakhtenko_random_1996}, generalized these results to band matrices,
where $\s_{\mu i,\nu j}$ is only required to be sufficiently smooth.  They found
\begin{align}
  \label{eq:CasatiGirko}
  \R_{\mu i}^t(z)=\frac{1}{z-a_{\mu i}-t\tG_{\mu i}^t(z)},
\end{align}
where
\begin{align}
 \label{eq:Gtilde}
  \tG_{\mu i}^t(z)\equiv\sum_{\nu j}\s_{\mu i,\nu j}^2 \R^t_{\nu j}
\end{align}
is the Wieltjes (short for weighted Stieltjes) transformation.  The Casati--Girko formalism requires a self-consistent solution of \eqref{eq:CasatiGirko} and \eqref{eq:Gtilde}.

Using this formalism, we can easily derive a formula~\cite{allez_eigenvector_2012,allez_eigenvector_2014,bourgade_eigenvector_2017,benigni_eigenvectors_2020}
for the overlap of the eigenvectors of $M^t$ vs.\ $A$.
We first apply the identity 
\begin{align}
  \label{eq:ppindentity}
  \frac{1}{\xi-\i0^+}=\mathcal{P}\Of(\frac{1}{\xi})+\i\pi\delta(\xi),
\end{align}
where $\mathcal{P}$ is the principal part and $0^+$ a positive infinitesimal, to \eqref{eq:R_diagonal},
\begin{align}
  \label{eq:Im_R_diagonal}
  \frac{1}{\pi}\Im R_{\mu i}^t\of(\xi-\i0^+)
  =\sum_n \bigabs{\braket{\phi_{\mu i}|\psi_n}}^2\delta\of(\xi-\l_n).
\end{align}
We then introduce
%
\begin{align}
  \Gamma_n &\equiv \intd_{-\infty}^{\frac{1}{2}(\l_n+\l_{n+1})}{\diff\xi}
  \frac{1}{\pi}\Im R_{\mu i}^t\of(\xi-\i0^+) \nonumber\\
  &=\sum_{m=1}^n \bigabs{\braket{\phi_{\mu i}|\psi_m}}^2,
  \label{eq:Gamma_n}
\end{align}
where we have assumed the eigenvalues of $M^t$ are ordered such that $\l_n<\l_{n+1}\ \forall\ n$.  For the statistical expectation values of the overlaps we find
\begin{align}
  \label{eq:chi_mu_i_n}
  \chi_{\mu i,n}
  &\equiv\E\Of[\bigabs{\braket{\phi_{\mu i}|\psi_n}}^2]
    =\E\Of[\Gamma_n-\Gamma_{n-1}]=\nonumber\\[5pt]
  &=\E\Of[\intd_{\frac{1}{2}(\l_{n-1}+\l_{n})}^{\frac{1}{2}(\l_n+\l_{n+1})}
    {\diff\xi}\frac{1}{\pi}\Im R_{\mu i}^t\of(\xi-\i0^+)]\nonumber\\[5pt]
  &=\E\Of[\frac{\l_{n-1}+\l_{n+1}}{2}]
    \frac{1}{\pi}\Im \R_{\mu i}^t\of(\l_n-\i0^+)\nonumber\\[5pt]
  &=\frac{1}{\pi N\rho(\l_n)}\Im \frac{1}{\l_n-a_{\mu i}-t\tG_{\mu i}^t(\l_n-\i0^+)},
\end{align}
where we have substituted \eqref{eq:CasatiGirko}. 
$\rho(\l)$ is the (normalized) eigenvalue density of $M^t$.

%
%
\section{Analytical results}
\label{sec:analytics}

In the context of our analysis, the diagonal matrix $A=\hH_\S+\hH_\B$ with eigenvalues $a_{\mu i}\equiv\eps_\mu+E_i$ describes the Hamiltonian of both system S and bath B in the absence of coupling, and $X^t$ is the coupling between the system and a small region R of the bath.  For the models we consider here, the variances for the random entities of $X^t$ factorize into dependencies on the system and the on the bath,
\begin{align}
  \label{eq:sigma}
  \s_{\mu i,\nu j}^2 =\s_{\mu\nu }^2\hspace{1pt} \t_{ij}^2.
\end{align}

If we assume that the distribution of the scattering elements in the matrix of the $N_\S N_\RR$ dimensional matrix $\hX^{\S\RR}$,
which acts only on the part of the Hilbert space describing S and R,
is uniform in the sense that its elements are not correlated with the energies of the states in R, $\t_{ij}^2$ takes the form
\begin{align}
  \label{eq:tau_ij}
  \t_{ij}^2= \frac{\e^{-b_0^{\,2}}}{N_\B\sqrt{\rho_\B(E_i)\rho_\B(E_j)}}
  \hspace{.5pt}\frac{1}{\spi\D_0}
  \exp\Of[-\frac{1}{\D_0^2}(E_i-E_j)^2],
\end{align}
where $\rho_\B(E)$ is the (normalized) eigenvalue density of the bath, $b_0\equiv\frac{1}{4}\b\D_0$, and $\D_0$ is a parameter which reflects that since $X^t$ only acts on degrees of freedom in the small region R of the bath B, it only scatters between states $\ket{\phi_{i}^B}$ and $\ket{\phi_{j}^B}$ which are accordingly close in energy.  In \appendixref{app:tau_variances}, we derive \eqref{eq:tau_ij} for an infinitesimal coupling between R and the remainder of the bath.  We find that
$\D_0=2\D_\RR$, where $\D_\RR$ is the variance of the eigenvalue density \eqref{eq:rho_R} of the region R.
Note that $\t_{ij}^2$ is normalized in the large bath limit,
\begin{align}
  \label{eq:tau_norm}
  \sum_{i} \t_{ij}^2
  &=N_\B\intd{\diff E}\rho_\B(E)\, \t_{ij}^2\Bigl|_{E_i\to E}\Bigr.
     =1,
\end{align}
where we have used $\rho_\B(E)=\rho_\B(E_j) \e^{\b(E-E_j)}$.
The form \eqref{eq:tau_ij} is consistent with our numerical work, and renders the problem amenable to analytic analysis.  The main result, the validity of the \ETH\ as outlined in Eqs.\ \eqref{eq:rho_n}--\eqref{eq:Z}, however, does not depend on the specific form we assume for $\t_{ij}^2$ (or $\s_{\mu\nu}^2$).

If the system S were sufficiently large such that only a region of it was coupled to the bath, we would obtain a form similar to \eqref{eq:tau_ij} for $\s_{\mu\nu }^2$.  In most of our numerical work, however, S consists only of two levels, and we obtain $\s_{\mu\nu }^2$ by inspection of the interaction.
%
%
%
%
%
With $\sum_{\mu}\hspace{-1pt}\s_{\mu\nu}^2 =1$,
the matrix variance of $X^t$ is
\begin{align}
  \label{eq:X_norm}
  \sum_{\mu i}\E\Of[\abs{X_{\mu i,\nu j}}^2]=t\quad\forall\,j.
\end{align}

We now assume that the overlaps \eqref{eq:chi_mu_i_n} 
will have the functional form
\begin{align}
  \label{eq:chi_mu_i_n_ansatz}
  \chi_{\mu i,n}=\frac{\chi_\mu(a_{\mu i}-\l_n-\eta_\mu)}
  {N\sqrt{\rho(a_{\mu i}-\eta_\mu)\rho(\l_n)}},
\end{align}
where $\rho$ is the (normalized) eigenvalue density of $M^t$.  The $\chi_\mu(x)$'s describe the peaks in Fig.~\ref{fig1}\figletter{a},
and the $\eta_\mu$'s are shift parameters.
We will justify this Ansatz \emph{a posteriori}.  Due to the completeness of the bases $\ket{\psi_n}$ and $\ket{\phi_{\mu i}}$, 
$\chi_\mu$ has to satisfy
\begin{align}
  1=\sum_n\chi_{\mu i,n}
  &=\intd{\diff\l} N\rho(\l)\,\chi_{\mu i,n}\Bigl|_{\l_n\to\l}\Bigr. \nonumber \\
  &=\intd{\diff x}\e^{-\half\b x}\chi_\mu(x)
  \label{eq:sum_n_chi}
\end{align}
and
\begin{align}
  1&=\sum_{\mu i}\chi_{\mu i,n} \nonumber\\
  &=\sum_\mu\intd{\diff a} N_\B\hspace{.5pt}\rho_\B(a-\eps_\mu)\,\chi_{\mu i,n}\Bigl|_{a_{\mu i}\to a}\Bigr.
  \label{eq:sum_mu_i_chi}
\end{align}
To evaluate \eqref{eq:sum_mu_i_chi}, we first relate the eigenvalue density of the bath $\rho_\B$ to the eigenvalue density $\rho_0$ of A,
\begin{align}
  \label{eq:rho0_vs_rho_B}
  \rho_0(a)=\frac{1}{N_\S}\sum_\mu\rho_\B(a-\eps_\mu)
  =\frac{\rho_\B(a)}{N_\S}Z_0^\S,
\end{align}
where $Z_0^\S=\sum_\mu\e^{-\b\eps_\mu}$ is the partition sum of the unperturbed subsystem S.  The difference between the eigenvalue densities of $A$ and $M^t$ may be parameterized through a shift $\eta$, \ie $\rho_0(a)=\rho(a-\eta)=\e^{-\b\eta}\rho(a)$.  This difference stems from the difference $t$ in the matrix variances of $A$ and $M^t$
\footnote{The reason we have to distinguish between $\rho_0$ and $\rho$ even though we assume the large bath limit is that as we take $\a\to 0^+$ in \eqref{eq:bn=-a_ln}, $\l,a\to -\infty$, while $\eta$ remains finite.}.
Substitution of \eqref{eq:chi_mu_i_n_ansatz} and \eqref{eq:rho0_vs_rho_B} into \eqref{eq:sum_mu_i_chi} yields
\begin{align}
  \label{eq:sum_mu_i_chi_2}
  1=\frac{1}{Z_0^\S}\sum_\mu\e^{-\b(\eps_\mu-\eta_\mu+\eta)}
  \intd{\diff x}\e^{+\half\b x}\chi_\mu(x).
\end{align}
If the $\chi_\mu(x)$'s are symmetric, the integral in \eqref{eq:sum_mu_i_chi_2} will be normalized to unity through \eqref{eq:sum_n_chi}.  In our numerics, we hardly observe any asymmetry, but this could change as we increase the coupling $t$.  In this case, the sensible approach is to define
the center of the peaks such that the two integrals remain equal, \ie
\begin{align}
  \label{eq:integral_unity}
  \intd{\diff x}\e^{+\half\b x}\chi_\mu(x)=1,
\end{align}
which leads to adjustments in the shifts $\eta_\mu$.
%
Note that \eqref{eq:sum_mu_i_chi_2} with \eqref{eq:integral_unity} fixes $\eta$.  For later purposes, we 
rewrite \eqref{eq:rho0_vs_rho_B} as
\begin{align}
  \label{eq:rho_B_vs_rho}
  \rho_\B(a)=\frac{N_\S}{Z^\S}\rho(a),
\end{align}
where $Z^\S =\sum_\mu\e^{-\b(\eps_\mu-\eta_\mu)}$ is the partition sum of the subsystem S in the presence of the perturbation $X^t$.

\vspace{\medskipamount}
%
%
To calculate $\chi_\mu(x)$,
we need to solve the Casati-Girko equations \eqref{eq:Gtilde} and \eqref{eq:CasatiGirko} self-consistently.  We do this iteratively, but operationally replace \eqref{eq:CasatiGirko} by \eqref{eq:chi_mu_i_n} and
the expectation value $\E[\,.\,]$ of \eqref{eq:R_diagonal}, \ie we do not map $\tG_{\mu i}^t(z)$ directly into $\R_{\mu i}^t(z)$, but first into $\chi_{\mu i,n}$, and then into $\R_{\mu i}^t(z)$.  The benefit is that we can begin the iteration with an Ansatz for 
$\chi_\mu(x)$.  We then calculate the imaginary part of $\R_{\mu i}^t(\xi-\i 0^+)$ via \eqref{eq:R_diagonal}, then the
imaginary part of $\tG_{\mu i}^t(\xi-\i 0^+)$ via \eqref{eq:Gtilde},
then the real part of $\tG_{\mu i}^t(\l_n)$ via Kramers-Kronig, and finally
$\chi_{\mu i,n}$ via \eqref{eq:chi_mu_i_n}.

Specifically, and in order to make the problem amenable to analytic calculations, we initially take a Gaussian for the peak function, 
\begin{align}
  \label{eq:chi_initial}
  \chi_\mu(x)=\frac{\e^{-b_\p^{\,2}}}{\spi \D_\p}
  \exp\Of{-\frac{x^2}{\D_\p^{\,2}}},
\end{align}
where $b_\p\equiv\frac{1}{4}\b \D_\p$.  Note that \eqref{eq:chi_initial} is normalized according to \eqref{eq:sum_n_chi} and \eqref{eq:integral_unity}.
After a single iteration, we will find that the peaks are described by Cauchy distributions with the half-widths $\gamma_\mu$ given in  \eqref{eq:intro_gamma_mu}, which for small $t$ are so narrow that, if we determine $\D_\p$ self consistently, it becomes negligible (compared to $\D_0$).
It is hence sufficient to assume a narrow peak for $\chi_\mu(x)$, and to carry out a single iteration.  Nothing will depend on the specific form of the peak.

We begin by calculating the imaginary part of $\R_{\nu j}^t$ 
for \eqref{eq:chi_initial} with \eqref{eq:R_diagonal} and \eqref{eq:ppindentity},
\begin{align}
  \label{eq:Im_R}
  \frac{1}{\pi}\Im\R_{\nu j}^t(\xi-\i 0^+)
  =\sum_n \E\Of[\bigabs{\braket{\phi_{\nu j}|\psi_n}}^2 \delta\of{\xi-\l_n}]
\end{align}
Since $\lim_{N\to\infty}\bigabs{\braket{\phi_{\nu j}|\psi_n}}^2\to\chi_{\nu j,n}$ due to self-averaging, we can evaluate the l.h.s.,
\begin{align}
  \label{eq:sum_for_Im_R}
  \sum_n &\chi_{\nu j,n} \,  \delta\of{\xi-\l_n}
  =\intd{\diff\l}
    N\rho(\l)\,\chi_{\nu j,n}\,\delta\of{\xi-\l_n}\Bigl|_{\l_n\to\l}\Bigr.
    \nonumber\\
  &=e^{-\half\b\of{a_{\nu j}-\eta_\nu -\xi}}\,
    \chi_\nu(a_{\nu j}-\xi-\eta_\nu)\nonumber\\
  &=\frac{1}{\spi \D_\p}
    \exp\Of[-\frac{1}{\D_\p^{\,2}}\Of{{a_{\nu j}-\eta_\nu -\xi}+b_\p \D_\p}^2].
\end{align}
From \eqref{eq:Gtilde} we obtain 
\begin{align}
  \label{eq:Im_Gtilde}
  &\frac{1}{\pi}\Im\tG_{\mu i}^t(\xi-\i 0^+)
  =\sum_\nu \s_{\mu\nu}^2 \sum_j \tau_{ij}^2\,
   \frac{1}{\pi}\Im\R_{\nu j}^t(\xi-\i 0^+) \nonumber\\
   &=\sum_\nu \s_{\mu\nu}^2 \hspace{1pt}\frac{1}{\pi\D_0\D_\p}\intd{\diff a}
\exp\Of[-\frac{1}{\D_\p^{\,2}}\Of{{a-\eta_\nu -\xi}+b_\p \D_\p}^2]
                    \nonumber\\
    &\hspace{50pt}\cdot\exp\Of[-\frac{1}{\D_0^{\,2}} \Of{(a-\eps_\nu)-(a_{\mu i}-\eps_\mu)-b_0\D_0}^2]
      \nonumber\\
  &=\sum_\nu \s_{\mu\nu}^2 \hspace{1pt}\frac{1}{\spi\D_1}
    \exp\Of[-\frac{1}{\D_1^{\,2}}
    \Of{a_{\mu i}-\xi-\eta_\mu-\eps_{\mu\nu}+b_1\D_1}^2],
\end{align}
where 
$\eps_{\mu\nu}\equiv(\eps_\mu-\eta_\mu)-(\eps_\nu-\eta_\nu)$, $\D_1^2\equiv\D_0^2+\D_\p^2$ and $b_1\equiv\frac{1}{4}\b \D_1$.

If we substitute $\E[\,.\,]$ of \eqref{eq:R_diagonal} into \eqref{eq:Gtilde}, we see that the real and imaginary part of $\tG_{\mu i}^t(z)$ are related by 
\begin{align}
  \label{eq:KK}
  \Re\tG_{\mu i}^t(\l_n)=\dashint\hspace{-2pt}\diff\xi\hspace{3pt}
  \frac{\frac{1}{\pi}\Im\tG_{\mu i}^t(\xi-\i 0^+)}{\l_n-\xi },
\end{align}
where $\dashint$ denotes the principle part $\mathcal{P}$ 
of the integral.
With \eqref{eq:Im_Gtilde} and
$u\equiv\frac{1}{\D_1}(\l_n-\xi)$,
$v\equiv\frac{1}{\D_1}(a_{\mu i}-\l_n-\eta_\mu-\eps_{\mu\nu}+b_1\D_1)$,
we obtain
\begin{align}
  \label{eq:Re_Gtilde}
  \Re\tG_{\mu i}^t(\l_n)
  =\frac{1}{\spi\D_1}\sum_\nu \s_{\mu\nu}^2\hspace{1pt}
  \dashint\hspace{-2pt}\diff u\hspace{3pt}\frac{\e^{-(u+v)^2}}{u}.
\end{align}
With
\begin{align}
  \label{eq:KK_integral}
  \dashint\hspace{-2pt}\diff u\hspace{3pt}\frac{\e^{-(u+v)^2}}{u}
  &=-\e^{-v^2}\sum_{m=0}^\infty \frac{(2v)^{2m+1}}{(2m+1)!}\hspace{1pt}
  \Gamma\Of{m+\half} \nonumber \\
  &=-\pi\e^{-v^2}\erfi(v) 
    \nonumber\\[3pt]
  &=-2\spi\Of{v-\frac{2v^3}{3}+\frac{4v^5}{15}-\ldots},
\end{align}
we obtain the Wieltjes transformation
\begin{align}
  \label{eq:Wieltjes}
  \tG_{\mu i}^t(\l_n-\i0^+)
  =\frac{\spi}{\D_1}\sum_\nu\s_{\mu\nu}^2\e^{-v^2}\Of{-\erfi(v)+\i\hspace{1pt}}.
\end{align}

To obtain the overlaps functions $\chi_\mu(x)$,
we first substitute \eqref{eq:chi_mu_i_n} into \eqref{eq:chi_mu_i_n_ansatz}.
This yields
\begin{align}
  \label{eq:chi_mu_form}
  &\chi_\mu(x)
  =\frac{1}{\pi}\e^{\half\b x}
  \Im\Of[ \frac{-1\hspace{10pt}}{x+\eta_\mu+t\tG_{\mu i}^t(\l_n-\i0^+)}]
    \nonumber\\[5pt]
  &=\frac{1}{\pi}
    \frac{t \e^{\half\b x} \Im\tG_{\mu i}^t(\l_n-\i0^+)\hspace{10pt}}
    {\Of{x+\eta_\mu+t\Re\tG_{\mu i}^t(\l_n)}^2 + \Of{t\Im\tG_{\mu i}^t(\l_n-\i0^+) }^2},
\end{align}
where we have defined $x\equiv a_{\mu i}-\l_n-\eta_\mu$.  There are two important results here.  The first is that since $\tG_{\mu i}^t(\l_n)$ depends on $a_{\mu i}$ and $\l_n$ only through $x$ in $v=\frac{1}{\D_1}(x-\eps_{\mu\nu})+b_1$, \eqref{eq:chi_mu_form} justifies the validity of the assumption \eqref{eq:chi_mu_i_n_ansatz}.  The second is that in the weak coupling limit, $\chi_\mu(x)$ is given to a highly accurate approximation by a
Cauchy distribution,
\begin{align}
  \label{eq:chi_mu_Lorentzian}
   \chi_\mu(x)\approx \frac{1}{\pi}\frac{\g_\mu}{x^2+\g_\mu^2},
\end{align}
and the lowest order contributions to both $\g_\mu$ and $\eta_\mu$ will be linear in $t$.  With \eqref{eq:Wieltjes}, we can easily derive \eqref{eq:intro_gamma_mu} and \eqref{eq:intro_eta_mu}.  Note that the numerator of \eqref{eq:chi_mu_form} effectively implements a (Gaussian) cutoff at $\abs{x}\sim\D_1$, which is insignificant since
$\g_{\mu}\ll\D_1 $
for small $t$.  This inequality further implies that if we adjust the width $\D_\p$ of \eqref{eq:chi_initial} to match $\g_\mu$ of \eqref{eq:chi_mu_Lorentzian},
we see that the difference between $\D_1$ and $\D_0$ is of
$\O\of(\sfrac{t^2}{\D_0^2})$.

It is possible to calculate $\g_\mu$ and $\eta_\mu$ to higher orders, but since
they are elaborate and
hardly visible in our numerical experiments,
we will not discuss the results here.

\vspace{\medskipamount}
%
%
We now proceed to calculate the reduced density matrix of the subsystem S. With \eqref{eq:rho_n}, we write the matrix elements
\begin{align}
  \label{eq:rho_mu_nu_def}
  \bra{\phi_\mu^\S}\hat\rho^\S_{n}\ket{\phi_\nu^\S}
  =\sum_i \braket{\phi_{\mu i}|\psi_n}\braket{\psi_n|\phi_{\nu i}}.
\end{align}
%
%
%
For the diagonal elements we obtain with
\eqref{eq:chi_mu_i_n_ansatz}, \eqref{eq:rho_B_vs_rho}, and \eqref{eq:integral_unity}
\begin{align}
  \label{eq:rho_mu_mu}
  \E\Big[\bra{\phi_\mu^\S}&\hat\rho^\S_{n}\ket{\phi_\mu^\S}\Big]
  =\sum_i \chi_{\mu i,n} \nonumber \\
    &=\intd{\diff a}
    N_\B\hspace{.5pt}\rho_\B(a-\eps_\mu)
    \frac{\chi_\mu(a-\l_n-\eta_\mu)}{N\sqrt{\rho(a-\eta_\mu)\rho(\l_n)}}
    \nonumber\\
  &=\frac{1}{Z^\S} \e^{-\b(\eps_\mu-\eta_\mu)}.
\end{align}
The diagonal elements hence describe a canonical ensemble with shifted energy levels $\eps_\mu-\eta_\mu$.  Note that the form of the peaks \eqref{eq:chi_mu_Lorentzian} does not enter here.  The canonical distribution follows exclusively from the validity of Ansatz \eqref{eq:chi_mu_i_n_ansatz} for $\chi_\mu\of{x}$ and the equivalence of the exponentially weighted areas \eqref{eq:sum_n_chi} and \eqref{eq:integral_unity} under $\chi_\mu\of{x}$ for all values of $\mu$, which in turn is due to the completeness of the basis states.  To justify \eqref{eq:chi_mu_i_n_ansatz}, however, we had to solve Casati--Girko, or more precisely, to demonstrate that $\tG_{\mu i}^t(\l_n)$ as given in \eqref{eq:Wieltjes}, and therefore $\chi_\mu(x)$ as given in \eqref{eq:chi_mu_form}, depends only through the variable $x$ on $a_{\mu i}$, $\l_n$ and $\eta_\mu$.
The approximation \eqref{eq:chi_mu_Lorentzian} with \eqref{eq:intro_gamma_mu} and \eqref{eq:intro_eta_mu} illustrates this general result and allows for comparison with our numerical results.

For the off-diagonal elements, we approximate the r.h.s.\ of \eqref{eq:rho_mu_nu_def} by a sum of uncorrelated, random numbers in the complex plane, which is equivalent to a random walk in 2D.  For the squares of the matrix elements, we find with
\eqref{eq:chi_mu_i_n_ansatz}, \eqref{eq:rho_B_vs_rho}, and \eqref{eq:chi_mu_Lorentzian}
\begin{align}
  \label{eq:rho_mu_nu_squared}
  &\E\Of[\abs{\bra{\phi_\mu^\S}\hat\rho^\S_{n}\ket{\phi_\nu^\S}}^2]
  =\sum_i \chi_{\mu i,n}\hspace{.5pt}\chi_{\nu i,n}
    \nonumber\\[5pt]
  &=\intd{\diff a} N_\B\hspace{.5pt} \rho_\B\Of(a-\eps_\mu)
    \frac{\chi_\mu\Of{a-\eta_\mu-\l_n}
     \chi_\nu\Of{a-\eta_\mu-\eps_{\mu\nu}-\l_n}}
    {N^2\rho\of(\l_n)\sqrt{\rho\of(a-\eta_\mu) \rho\of(a-\eta_\mu-\eps_{\mu\nu})}}
    \nonumber\\[5pt]
  &=\frac{\e^{-\half\b((\eps_\mu-\eta_\mu)+(\eps_\nu-\eta_\nu))}}{N\rho\of(\l_n)Z^\S}
    \intd{\diff x}\chi_\mu\Of{x} \chi_\nu\Of{x-\eps_{\mu\nu}}
    \nonumber\\[5pt]
  &=\frac{\e^{-\half\b((\eps_\mu-\eta_\mu)+(\eps_\nu-\eta_\nu))}}{N\rho\of(\l_n)Z^\S}
    \frac{1}{\pi}\frac{\g_\mu+\g_\nu}{\eps_{\mu\nu}^{\,2}+(\g_\mu+\g_\nu)^2}.
\end{align}
Since the dimension $N$ of the Hilbert space of the bath increases exponentially with the size of the bath, the off-diagonal entities of $\hat\rho^\S_{n}$ vanish very rapidly in the large bath limit, $\left(\hat\rho^\S_{n}\right)_{\text{od}}\propto \sfrac{1}{\sqrt{N}}\propto \sfrac{1}{\sqrt{N_\B}}$.  This result is consistent with our numerics (see Fig.~\ref{fig3}\figletter{d}).  It is also consistent with results by Beugeling, Moessner, and Haque~\cite{beugeling_finite-size_2014,beugeling_off-diagonal_2015}, who observed numerically that eigenstate-to-eigenstate fluctuations of typical observables in the context of eigenstate thermalization scale as $\sfrac{1}{\sqrt{N}}$.

\vspace{\medskipamount}
%
%
 Finally, we calculate the scattering matrix elements when we induce a transition between states $\ket{\phi_\mu}$ and $\ket{\phi_\nu}$ by externally perturbing the subsystem S, 
\begin{align}
  \label{eq:17}
  \mathcal{S}_{\mu\to\nu}(\o)
  &\equiv\bra{\psi_m}
  \bigl(\ket{\phi_\nu^\S}\bra{\phi_\mu^\S}\otimes\1^\B\bigr)
  \ket{\psi_n}\big|_{\l_m=\l_n+\o}\bigr.
    \nonumber\\[5pt]
  &=\sum_i \braket{\psi_m|\phi_{\nu i}}\braket{\phi_{\mu i}|\psi_n}
    \big|_{\l_m=\l_n+\o}\bigr.\, .
\end{align}
A calculation very similar to \eqref{eq:rho_mu_nu_squared} yields
\begin{align}
  \label{eq:S_mu_nu_squared}
  &\E\Of[\abs{\mathcal{S}_{\mu\to\nu}(\o)}^2]
  =\sum_i \chi_{\mu i,n}\hspace{.5pt}\chi_{\nu i,m}\big|_{\l_m=\l_n+\o}\bigr.
    \nonumber\\[5pt]
  &=\intd{\diff a}
    \frac{N_\B\hspace{.5pt} \rho_\B\Of(a-\eps_\mu)\chi_\mu\Of{a-\eta_\mu-\l_n}}
    {N^2\sqrt{\rho\of(\l_n)\rho\of(a-\eta_\mu)}} \nonumber\\[5pt]
    &\hspace{33pt}\cdot \frac{\chi_\nu\Of{a-\eta_\mu-\eps_{\mu\nu}-\l_n-\o}}{\sqrt{\rho\of(\l_n+\o)\rho\of(a-\eta_\mu-\eps_{\mu\nu})}}
    \nonumber\\[5pt]
 &=\frac{\e^{-\half\b((\eps_\mu-\eta_\mu)+(\eps_\nu-\eta_\nu))}}
   {N\sqrt{\rho\of(\l_n)\rho\of(\l_n+\o)}}\frac{1}{Z^\S}
   \frac{1}{\pi}\frac{\g_\mu+\g_\nu}{(\o+\eps_{\mu\nu})^2+(\g_\mu+\g_\nu)^2}.
\end{align}
The individual matrix element squares vanish again with the level spacing
$1/N\rho\of(\l)$ as we enlarge the bath, but the integral over any given 
energy interval will remain unaffected.  The important conclusion from formula \eqref{eq:S_mu_nu_squared} is that the peaks occur at $\o=-\eps_{\mu\nu}=(\eps_\nu-\eta_\nu)-(\eps_\mu-\eta_\mu)$, \ie we observe the shifted energy levels, which also enter the Boltzmann factors in \eqref{eq:rho_mu_mu}.  This shift, as well as the Lorentzian broadening to half-width $\g_\mu+\g_\nu$, are observable in our numerics and could conceivably be observed in an experimental realization with cold atoms.

\section{General applicability of the theory}
\label{sec:generalization}

%
%
So far, we have considered weak coupling matrices $\hX$ with random interactions between a small subsystems S and a large bath B. The assumption of weak couplings makes the systems amenable to analytical solution.  If $t$ is not small compared to $\D_0=2\D_\RR$, the iterative solution of the Casati--Girko formalism becomes more elaborate in several regards.  First, we have to expand $\g_\mu$ and $\eta_\mu$ to higher orders in $t/\D_0$.  Second, we will find deviations from the Cauchy distribution \eqref{eq:chi_mu_Lorentzian}, and the resulting curves can no longer be fully described by a single parameter $\g_\mu$.  Third, we can no longer solve Casati--Girko reliably with a single iteration, and have to replace \eqref{eq:chi_initial} in subsequent iterations with the result \eqref{eq:chi_mu_form} of the previous iteration.  The convolution of two Gaussians in \eqref{eq:Im_Gtilde} will be replaced approximately by a Voigt profile, which can no longer be evaluated analytically.

Most importantly, however, our Ansatz \eqref{eq:chi_mu_i_n_ansatz} will still be valid, as detailed in \appendixref{app:validity}.  Therefore, the reduced density matrix of the subsystem S will still be given the canonical distribution \eqref{eq:rho_mu_mu}, with off-diagonal elements \eqref{eq:rho_mu_nu_squared} which vanish exponentially as the size of the bath is increased.  The central result of this work is hence not limited to weak couplings, but generally valid.

The other assumption we have made, the assumption of random couplings, appears more problematic.  Recall the setup for our numerical work. Suppose we have a very strong onsite magnetic field acting on a spin $\hs_i$ in the region R of the bath B, which is strongly coupled (\eg via a Heisenberg term) 
to the single spin $\hs_i$ of the subsystem S, and that $\hH_\S$ is negligible by comparison.  In this case, the assumption of random couplings, and hence our analysis, is no longer applicable.
According to the \ETH , the subsystem S will still thermalize, but the reduced density matrix
$\hat\rho^\S$ will be diagonal in a basis determined by the mean field induced by the interaction
rather than that of $\hH_\S$.  At first sight, it appears as our theory of eigenstate thermalization was limited to random couplings, while the \ETH\ is not.


We can, however, easily generalize our derivation.
Since we assume a very large bath B and local interactions, our analysis will apply rigorously to any large intermediate subsystem M, which contains S and is small compared to B, but still sufficiently large such that the interactions between M and B can be taken random in the sense of self averaging due to concentration of measure.  Then our analysis shows that if the entire system is in an energy eigenstate, the reduced density matrix of M will be thermal with the temperature fixed by \eqref{eq:Hb=ln} with \eqref{eq:Z}  (see \appendixref{app:large_system_S} for an evaluation of $\g$ and $\eta$ for a larger subsystem with a continuous eigenvalue density).  This implies that any subsystem S of M will be in a thermal state, or more precisely, be described by the thermal density matrix \eqref{eq:rho_beta_n} with this temperature.

Therefore, our theory of eigenstate thermalization validates the hypothesis \eqref{eq:rho_n}--\eqref{eq:Z} of Deutsch and Srednicki in general.

\section*{Author contributions}

MG initiated and supervised the project.  THe and THo performed the numerical work.  MG, with THe and THo, worked out the analytical calculations.  All authors contributed to the work and the presentation, with THe and THo creating the figures and MG writing most of the text.

\section*{Acknowledgements}

We wish to thank Ludwig Bordfeldt, David Huse, Tom Oeffner, and Mark Srednicki for helpful suggestions on the manuscript. This work was supported by the Deutsche Forschungsgemeinschaft (DFG, German Research Foundation)---Project-ID 258499086---SFB 1170, through the Würz\-burg--Dresden Cluster of Excellence on Complexity and Topology in Quantum Matter---\textit{ct.qmat} Project-ID 390858490---EXC 2147, through DFG grant GR-1715/3-1 (MG), and through the German Academic Scholarship Foundation (THe).

%
%

\appendix

\begin{figure*}[!t]
  \centering
  \includegraphics[width=0.92\linewidth]
  {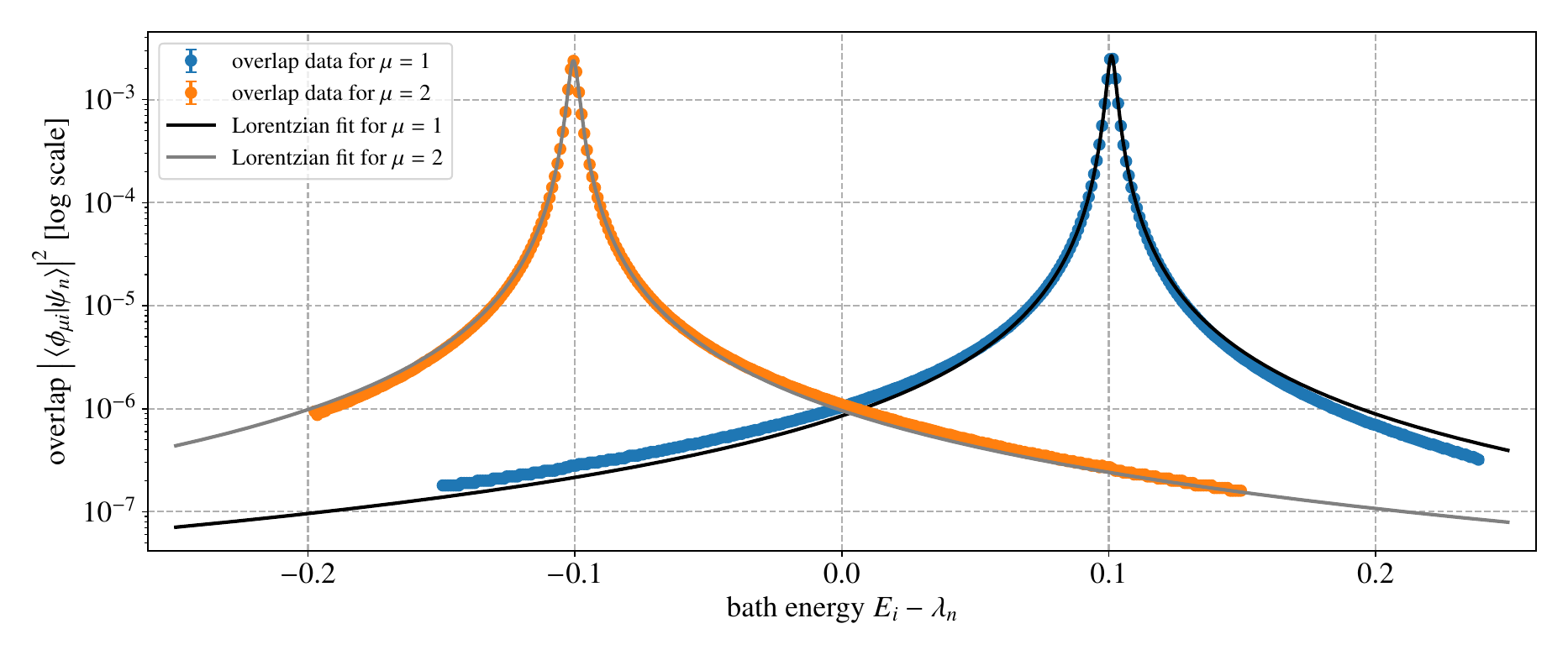}
  \caption{
    \label{fig:Lorentzians_log}
    Numerical data of the absolute squares of the overlaps $\abs{\!\braket{\phi_{\mu i}|\psi_n}\!}^2$ plotted for $\mu=1$ (blue) and $\mu=2$ (orange) on a logarithmic scale. The data are the same as depicted in Fig.~\ref{fig1}\figletter{b} and were obtained in an 18 site lattice with $t=\num{6.25e-4}$ by averaging over $N_\text{av} = 3200$ states $\ket{\psi_n}$ around the energy $\l_n = \num{-0.95789 \pm 0.00025}$.
    Close to their peak, the overlap data agrees with the Lorentzian fits (black for $\mu=1$; grey for $\mu=2$).  We observe small deviations and asymmetries far away from the peak positions.
  }
\end{figure*}

\section{Nesting of the \ETH}
\label{app:nesting}

\begin{figure*}[!t]
  \centering
  \includegraphics[width=0.92\linewidth]
  {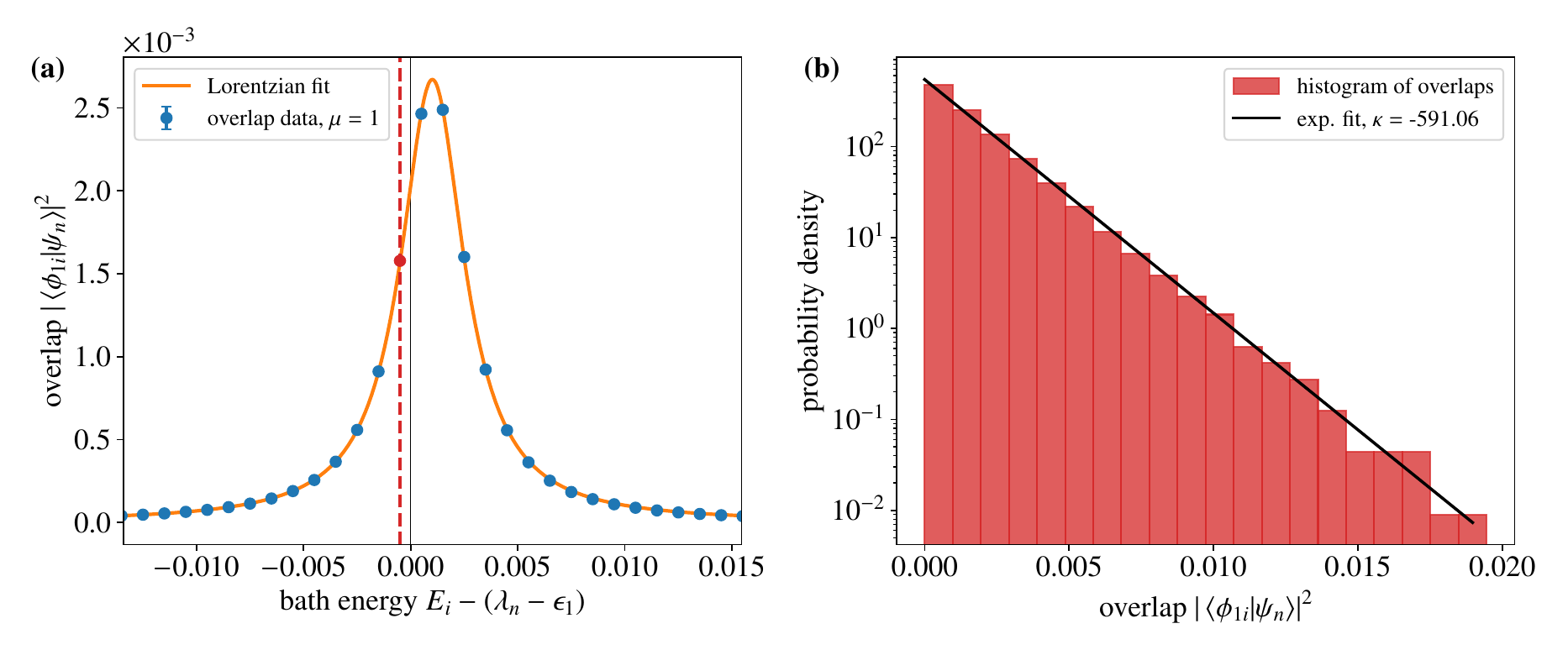}
  \caption{
    \label{fig:Lorentzians_statistics}
    \textbf{(a)} Numerical data of the absolute squares of the overlaps $\abs{\!\braket{\phi_{\mu i}|\psi_n}\!}^2$ for $\mu=1$ (blue) with a Lorentzian fit (orange).  We use the data shown in Fig.~\ref{fig1}\figletter{b} in blue.
    \textbf{(b)} Distribution of overlap values for one data point marked in red in (a) close to the left-handed half-width at half-maximum (HWHM) at $E_i - (\l_n - \eps_1 +\eta_1 ) \approx -\g_1$ on a logarithmic scale. An exponential fit confirms the exponential distribution of the squares of the overlaps.
  }
\end{figure*}

\begin{figure*}[!t]
  \centering
  \includegraphics[width=0.92\linewidth]
  {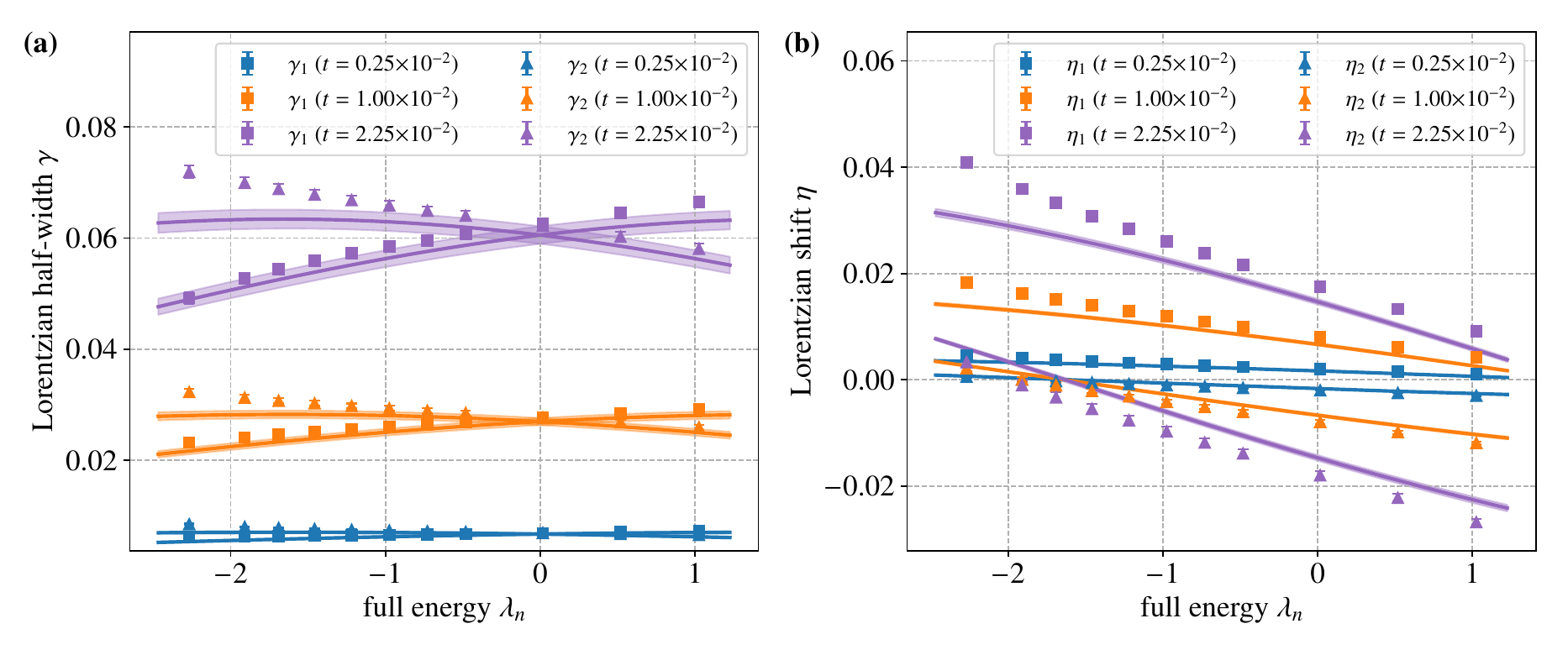}
  \caption{
    \label{fig:Lorentzian_params_vs_E}
    \textbf{(a)} Lorentzian half-widths $\g_{\mu,n}$ (\cf \eqref{eq:intro_gamma_mu}) for the two peaks of a system with 1 site in S and 15 sites in B, plotted as a function of the total energy $\l_n$
    \textbf{(b)} Shifts $\eta_{\mu,n}$ (\cf \eqref{eq:intro_eta_mu}) for the same system, again plotted as a function of $\l_n$.
    In both plots, the analytical results are shown as solid lines, with error intervals originating from the fit error of the parameters inserted into \eqref{eq:intro_gamma_mu} and \eqref{eq:intro_eta_mu}.
    In Fig.~\ref{fig3}\figletter{a} and \figletter{b}, we present the same data in terms of the sum and differences $\g_1 \pm \g_2$ and $\eta_1\pm\eta_2$.
  }
\end{figure*}

\begin{figure*}[!t]
  \centering
  \includegraphics[width=0.92\linewidth]
  {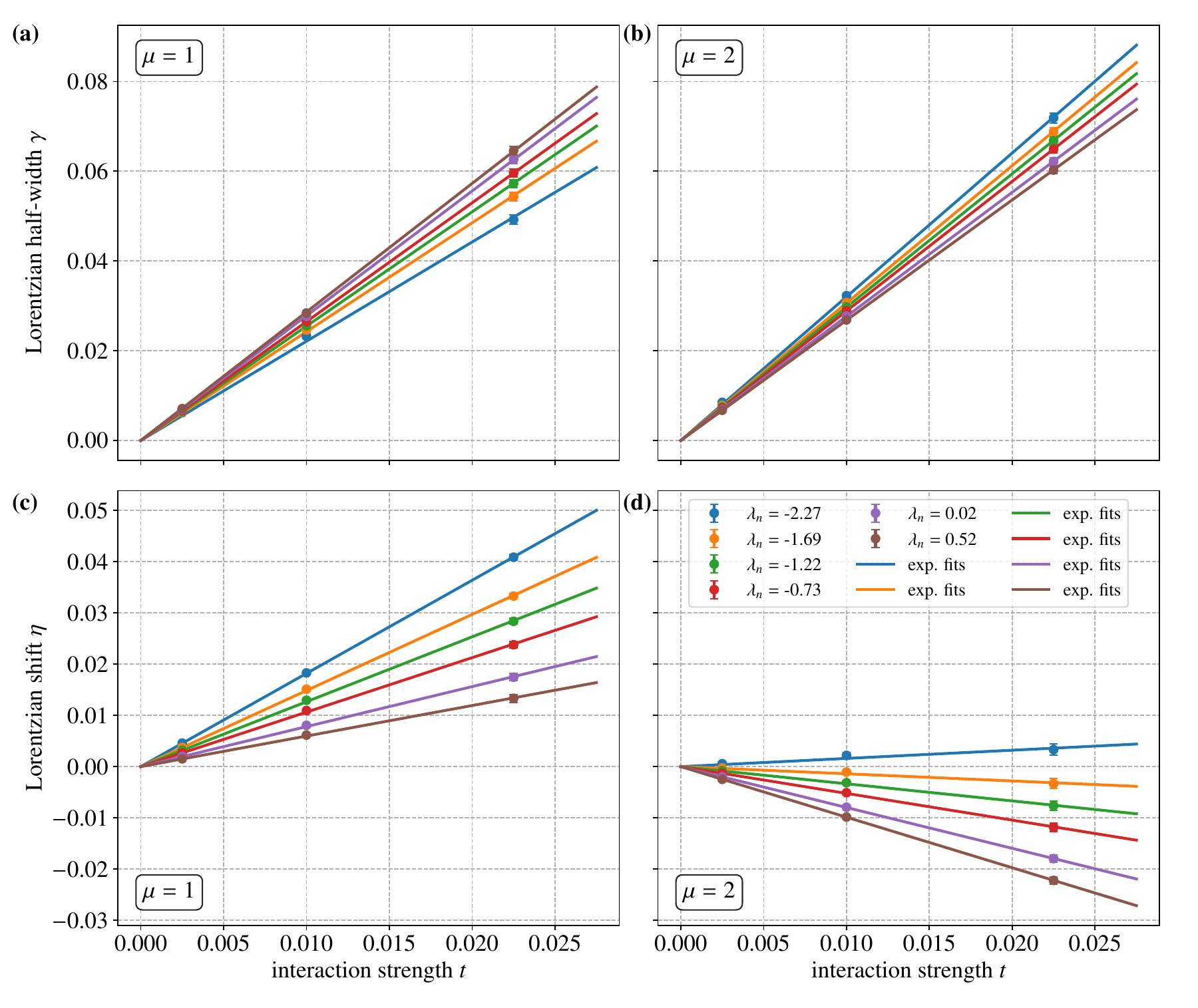}
  \caption{
    \label{fig:Lorentzian_params_vs_t}
    Lorentzian half-widths  \textbf{(a)} $\g_{1,n}$ and \textbf{(b)} $\g_{2,n}$ as well as shifts \textbf{(c)} $\eta_{1,n}$ and \textbf{(d)} $\eta_{2,n}$ plotted as functions of the interaction strength $t$, where $t$ is the second moment of the interaction matrix $\hX$.
    The different colors encode 
    different energies $\l_n$. Linear fits through the origin confirm the proportionality of all displayed values to $t$.
  }
\end{figure*}

\begin{figure*}[!t]
  \centering
  \includegraphics[width=0.95\linewidth]
  {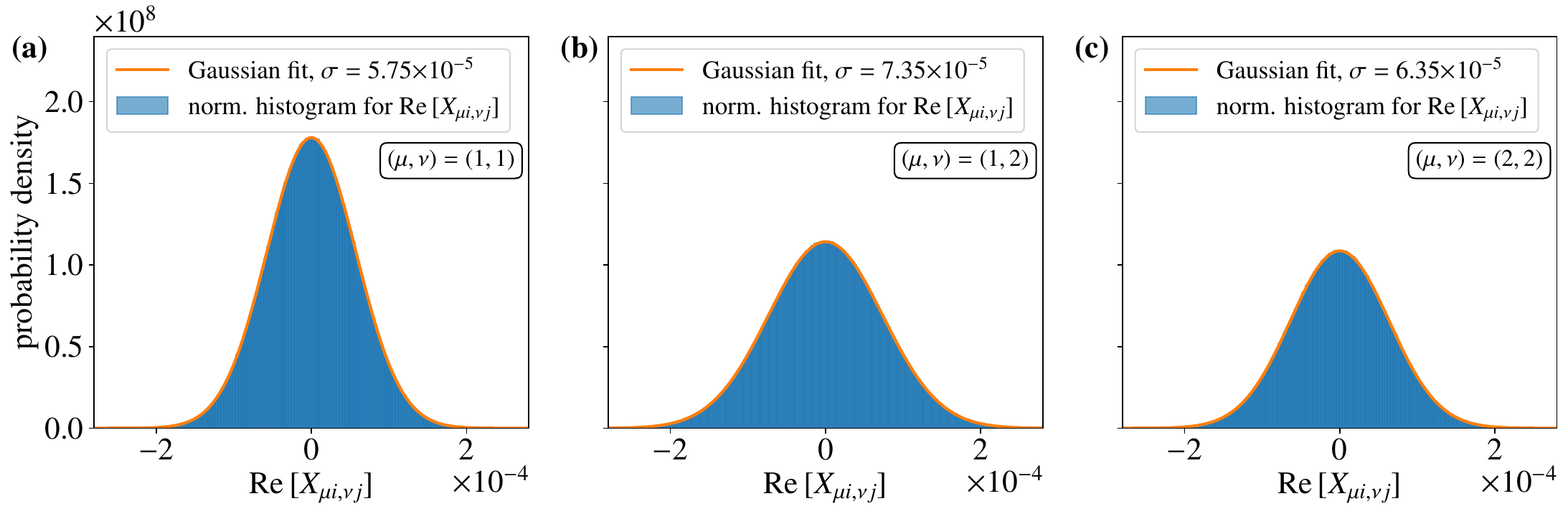}
  \caption{
    \label{fig:17+1sites_X_matrix_Re}
    Analysis of the off-diagonal elements $X_{\mu i, \nu j}$ of the interaction matrix $\hX$ between S and B with variance $t = \num{6.25e-4}$ for an 18 site lattice.  With eigenenergies $\eps_{1,2}=\mp 0.1$ in S, we extract boxes of width $\delta E_i = 0.1$ around the bath energies $(E_i, E_j) = (\lambda_n - \epsilon_\mu, \lambda_n - \epsilon_\nu)$ of the scattering $(\mu, \nu)$ in S, where $E_i$ is the row and $E_j$ the column energy.
    We choose $\l_n = \num{-0.95789 \pm 0.00025}$ in correspondence with Figs.~\ref{fig:Lorentzians_log} and~\ref{fig:Lorentzians_statistics}.
    This leads to boxes centered around the approximate location of the overlap peaks of $\abs{\braket{\phi_{\mu i}|\psi_n}}^2$ at $E_i \approx \l_n - \eps_\mu$ for $\mu = 1,2$.
    We show the distribution of the real part of the off-diagonal elements for the scatterings $(\mu,\nu)=$ \textbf{(a)} $(1, 1)$, \textbf{(b)} $(1, 2)$ and \textbf{(c)} $(2, 2)$. Gaussian fits in orange agree well with the numerical data, with their fitted standard deviation denoted in the legend of each plot.
  }
\end{figure*}

\begin{figure*}[!t]
  \centering
  \includegraphics[width=0.95\linewidth]
  {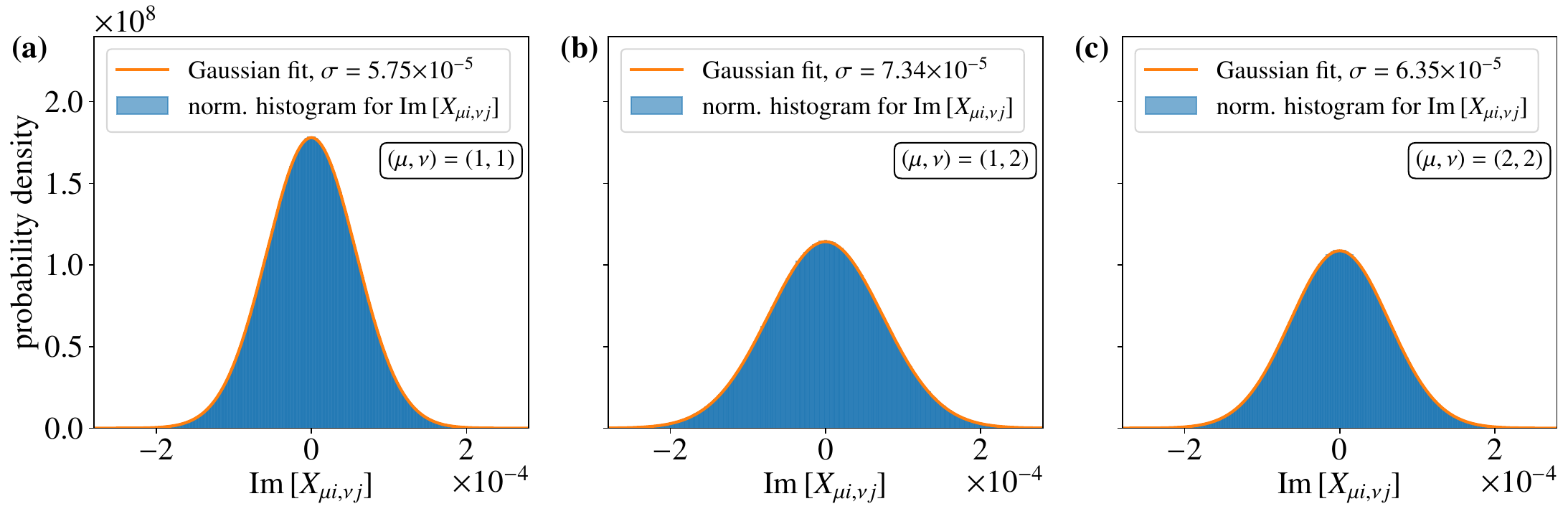}
  \caption{
    \label{fig:17+1sites_X_matrix_Im}
    Distribution of the imaginary  part of the off-diagonal elements for the scatterings shown in Fig.~\ref{fig:17+1sites_X_matrix_Re}.  The fitted standard deviations denoted in the legend of each plot are equal to the standard deviations of the real part shown there.
  }
\end{figure*}

\begin{figure*}[!t]
  \centering
  \includegraphics[width=0.95\linewidth]
  {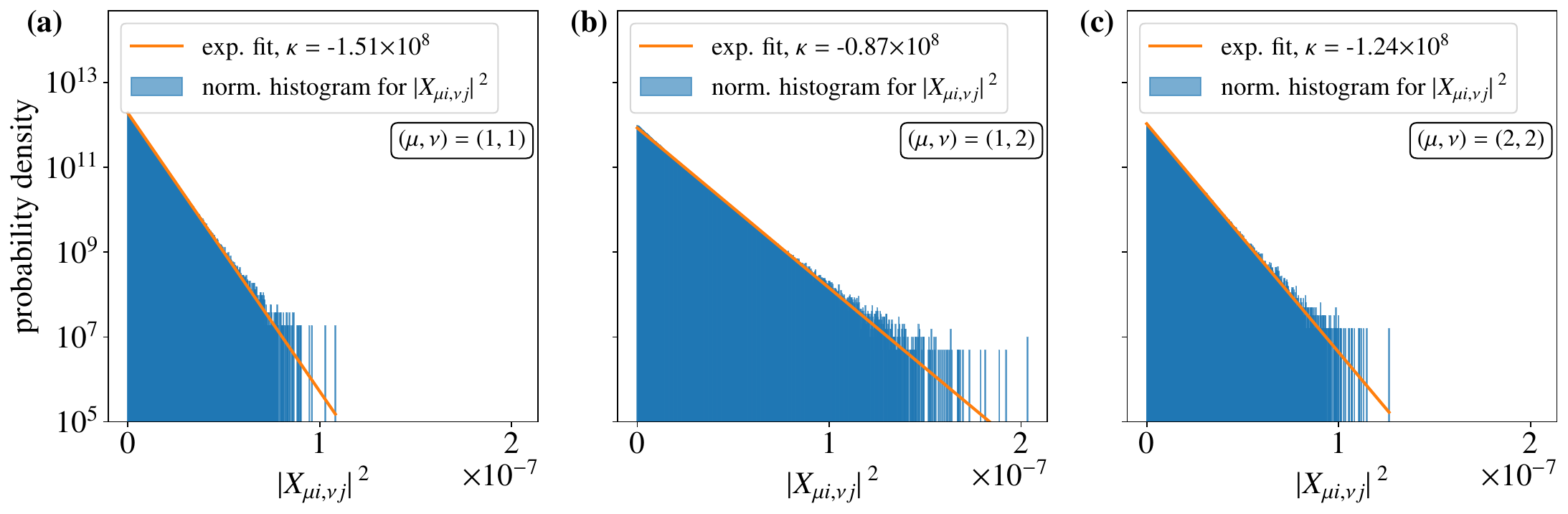}
  \caption{
    \label{fig:17+1sites_X_matrix_Abs2}
    Distributions  of the squares of the absolute value of the off-diagonal elements for the scatterings shown in Figs.~\ref{fig:17+1sites_X_matrix_Re} and~\ref{fig:17+1sites_X_matrix_Im}.
    Exponential fits in orange agree well with the numerical data, confirming an exponential probability distribution of the absolute squares of the matrix elements $\abs{X_{\mu i, \nu j}}^2$ due to the identical Gaussian distributions of the real and imaginary parts in Figs.~\ref{fig:17+1sites_X_matrix_Re} and~\ref{fig:17+1sites_X_matrix_Im}.
  }
\end{figure*}

\begin{figure*}[!t]
  \centering
  \includegraphics[width=0.9\linewidth]
  {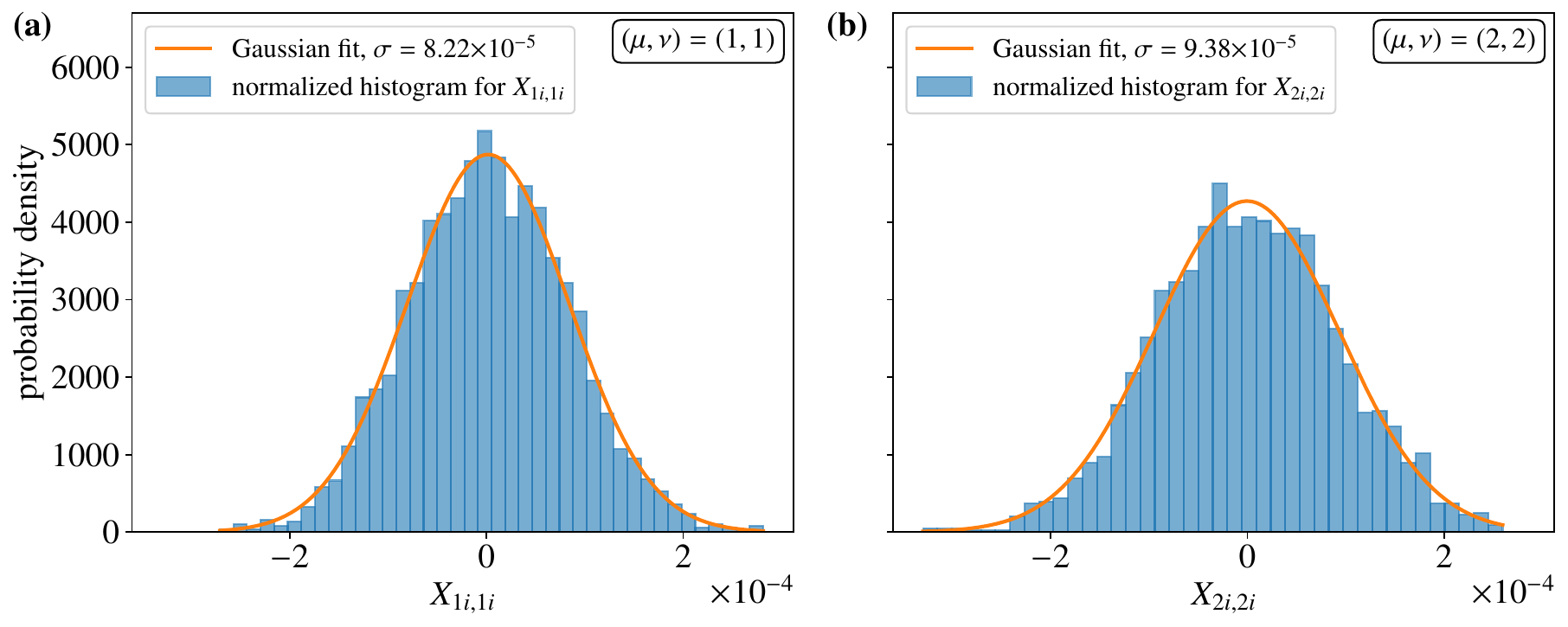}
  \caption{
    \label{fig:17+1sites_X_matrix_Dia}
    Analysis of the diagonal elements of the interaction matrix $\hX$ we analyzed in Fig.~\ref{fig:17+1sites_X_matrix_Re} and~\ref{fig:17+1sites_X_matrix_Im}.
    We again extract boxes of width $\delta E_i = 0.1$ around the bath energies $(E_i, E_i) = (\lambda_n - \epsilon_\mu, \lambda_n - \epsilon_\mu)$ for $\mu=1,2$.
    This leads to the boxes centered around the approximate location of the overlap peaks of $\abs{\braket{\phi_{\mu i}|\psi_n}}^2$ at $E_i \approx \l_n - \eps_\mu$ for $\mu = 1,2$. We show the distribution of the diagonal elements in the boxes for \textbf{(a)} $\mu=1$ and \textbf{(b)} $\mu=2$.
    Gaussian fits in orange agree well with the numerical data, while their fitted standard deviation denoted in the legends is $\sqrt{2}$ times the standard deviation of the real part for the corresponding plots with $\mu=\nu$ in Fig.~\ref{fig:17+1sites_X_matrix_Re}.
  }
\end{figure*}

\begin{figure*}[!t]
  \centering
  \includegraphics[width=0.9\linewidth]
  {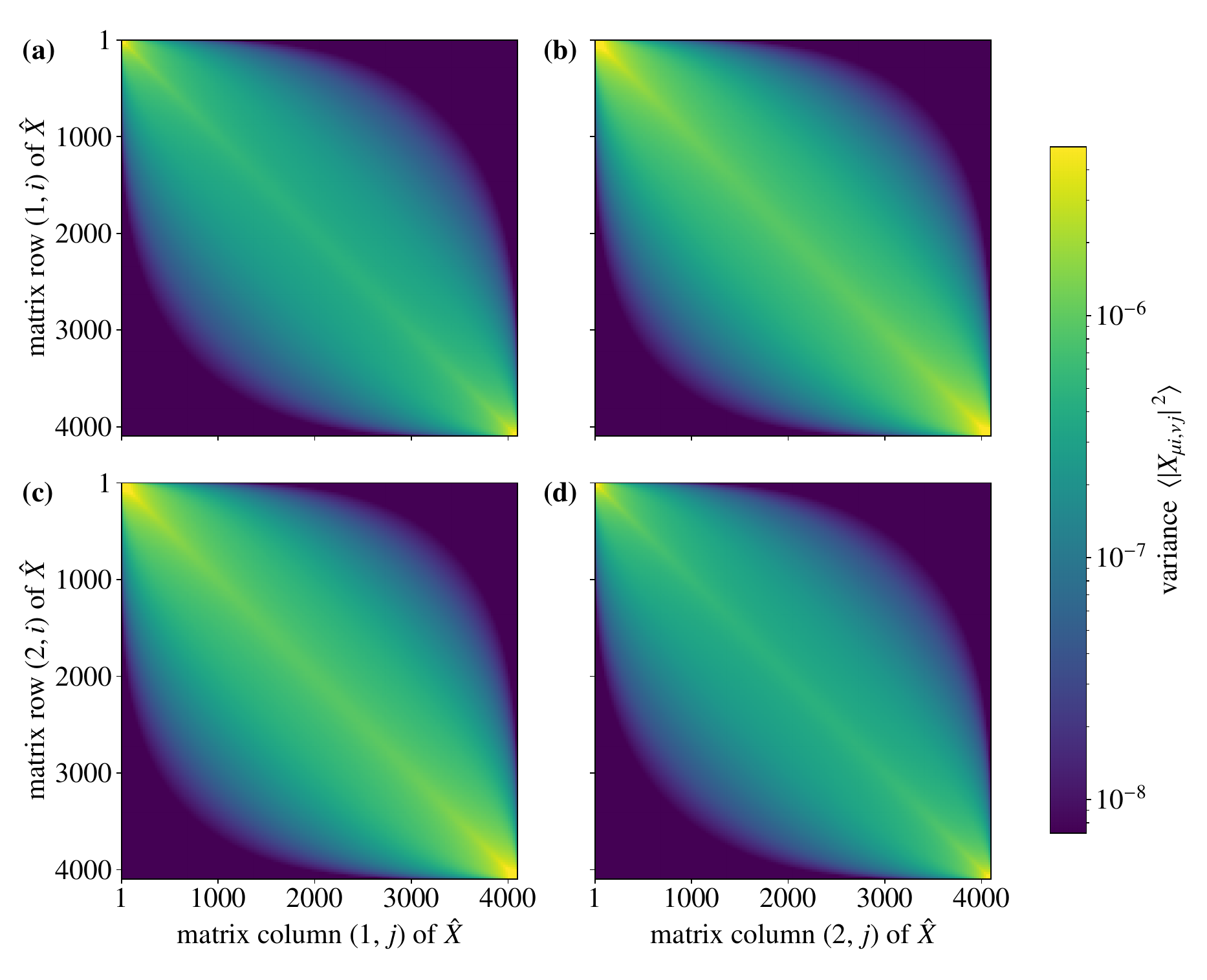}
  \caption{
    \label{fig:12+1sites_heatmap_log}
    Heat map representation with nearest shading type of the variance of the off-diagonal elements of the interaction $\hX$ with $t = \num{2.5e-3}$, obtained in boxes of size $(25 \times 25)$ matrix elements in a 13 site lattice and plotted with respect to the matrix indices. The variance is averaged over $100$ random samples of $\hH_\B$ and two samples of $\hX$ for each $\hH_\B$ and plotted on a logarithmic color scale. The matrix is separated into four blocks, where the substructure describes the scattering of eigenstates with energies $(E_i, E_j)$ in B.
    Each block corresponds to a scattering of eigenstates in S with $(\mu,\nu)=$ \textbf{(a)} $(1, 1)$, \textbf{(b)} $(1, 2)$, \textbf{(c)} $(2, 1)$ and \textbf{(d)} $(2, 2)$. We observe a ``leaf''-shaped distribution due to the suppressed eigenvalue density towards the boundary of the spectrum.
  }
\end{figure*}

\begin{figure*}[!t]
  \centering
  \includegraphics[width=0.9\linewidth]
  {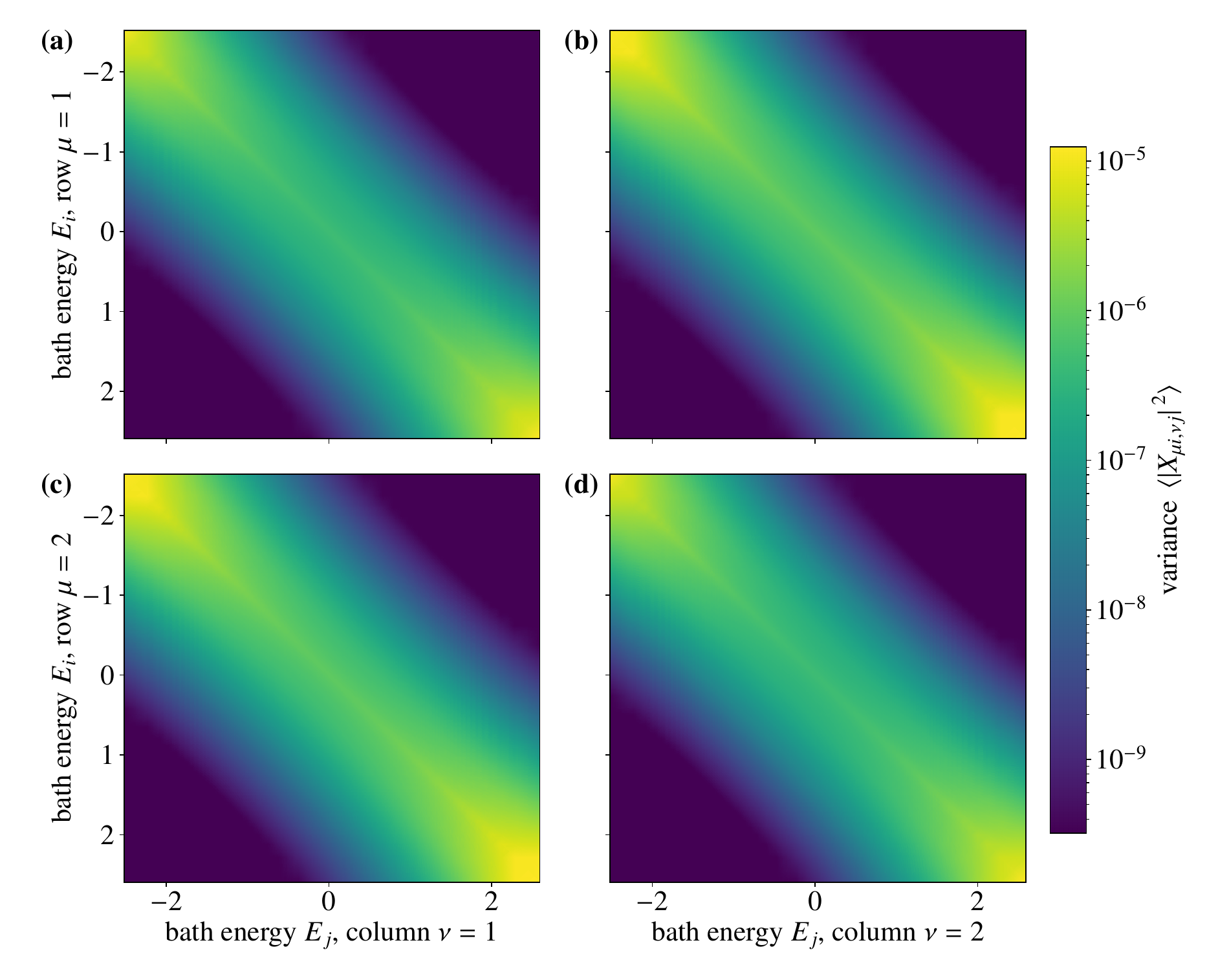}
  \caption{
    \label{fig:12+1sites_heatmap_energy_log}
    Heat map representation with Grioud shading type of the off-diagonal variance $\s_{\mu i,\nu j}^2$ of $\hX$ for the eigenstate scattering $(\mu,\nu)=$ \textbf{(a)} $(1, 1)$, \textbf{(b)} $(1, 2)$, \textbf{(c)} $(2, 1)$ and \textbf{(d)} $(2, 2)$ for the 13 site system of Fig.~\ref{fig:12+1sites_heatmap_log}.
    The variance is again obtained in boxes with $(25 \times 25)$ elements and plotted as a function of the mean bath energies $E_i$ and $E_j$.
    The variance along the diagonal direction of the matrix increases towards the edges, and falls off in the perpendicular direction. The band character of the matrix $\hX$ with a well-defined, energy-independent scattering width $\D_0$ is visible.
  }
  \end{figure*}

\begin{figure*}[!t]
  \centering
  \includegraphics[width=0.9\linewidth]
  {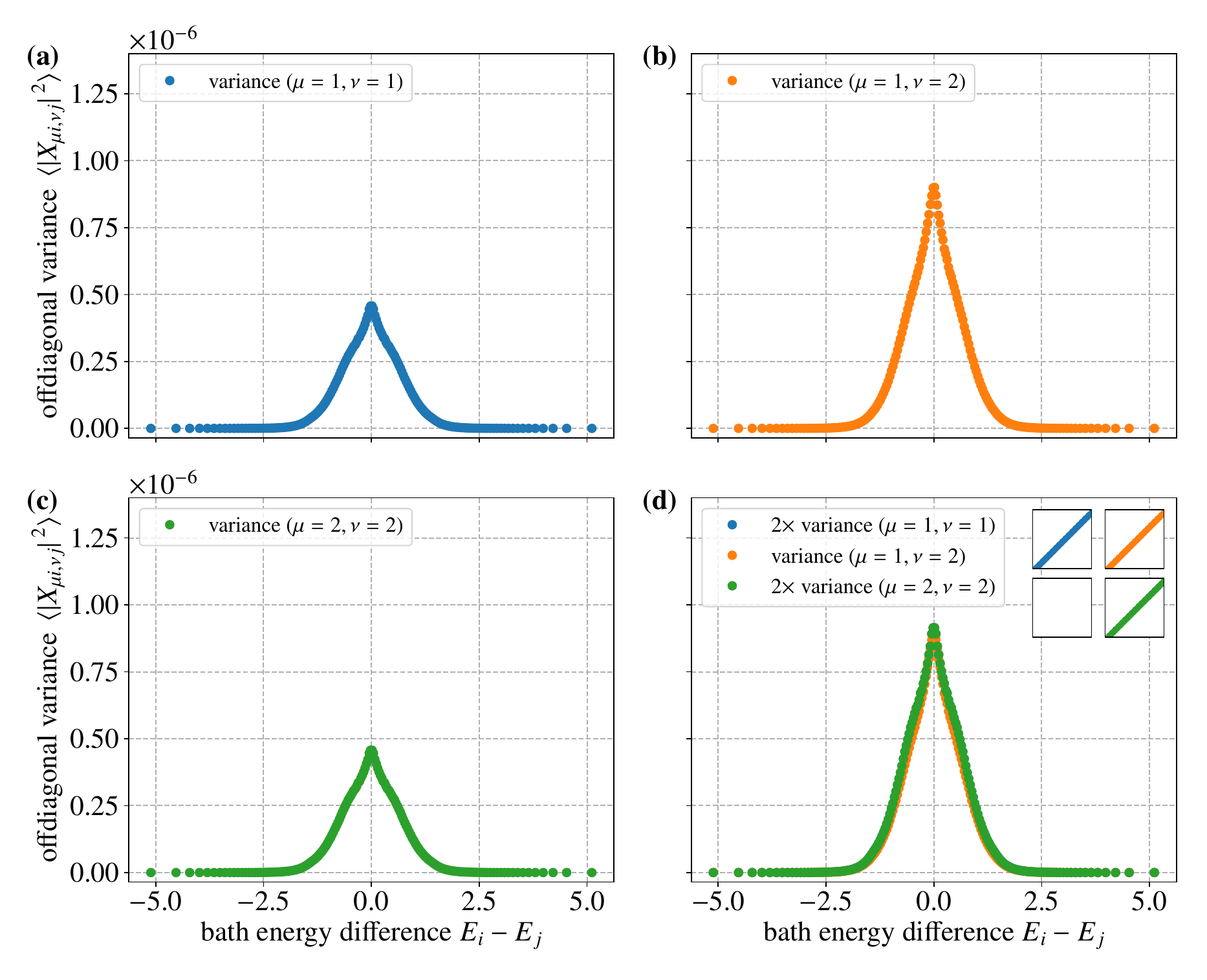}
  \caption{
    Off-diagonal variances $\s_{\mu i,\nu j}^2$ of $\hX$ in a 13 site lattice with $t = \num{2.5e-3}$ for the eigenstate scattering $(\mu,\nu)=$ \textbf{(a)} $(1, 1)$, \textbf{(b)} $(1, 2)$ and \textbf{(c)} $(2, 2)$.
    We plot the variance for a cross-section along the anti-diagonal of the matrix from the data of Fig.~\ref{fig:12+1sites_heatmap_energy_log} as a function of the bath energy difference $E_i-E_j$.  In \textbf{(d)}, we plot \textbf{(a-c)} on top of each other with the variances for $\mu=\nu$ scaled by a factor of two.
    The variance in all four blocks is identical except for a global prefactor, confirming the factorization $\s_{\mu i,\nu j}^2 =\s_{\mu\nu }^2\hspace{1pt} \t_{ij}^2$ assumed in~\eqref{eq:sigma} with $\s_{11}^2 = \s_{22}^2 = 1/3$ and $\s_{12}^2 = \s_{21}^2 = 2/3$.
    \label{fig:12+1sites_anti_dia}
  }
\end{figure*}

\begin{figure*}[!t]
  \centering
  \includegraphics[width=0.9\linewidth]
  {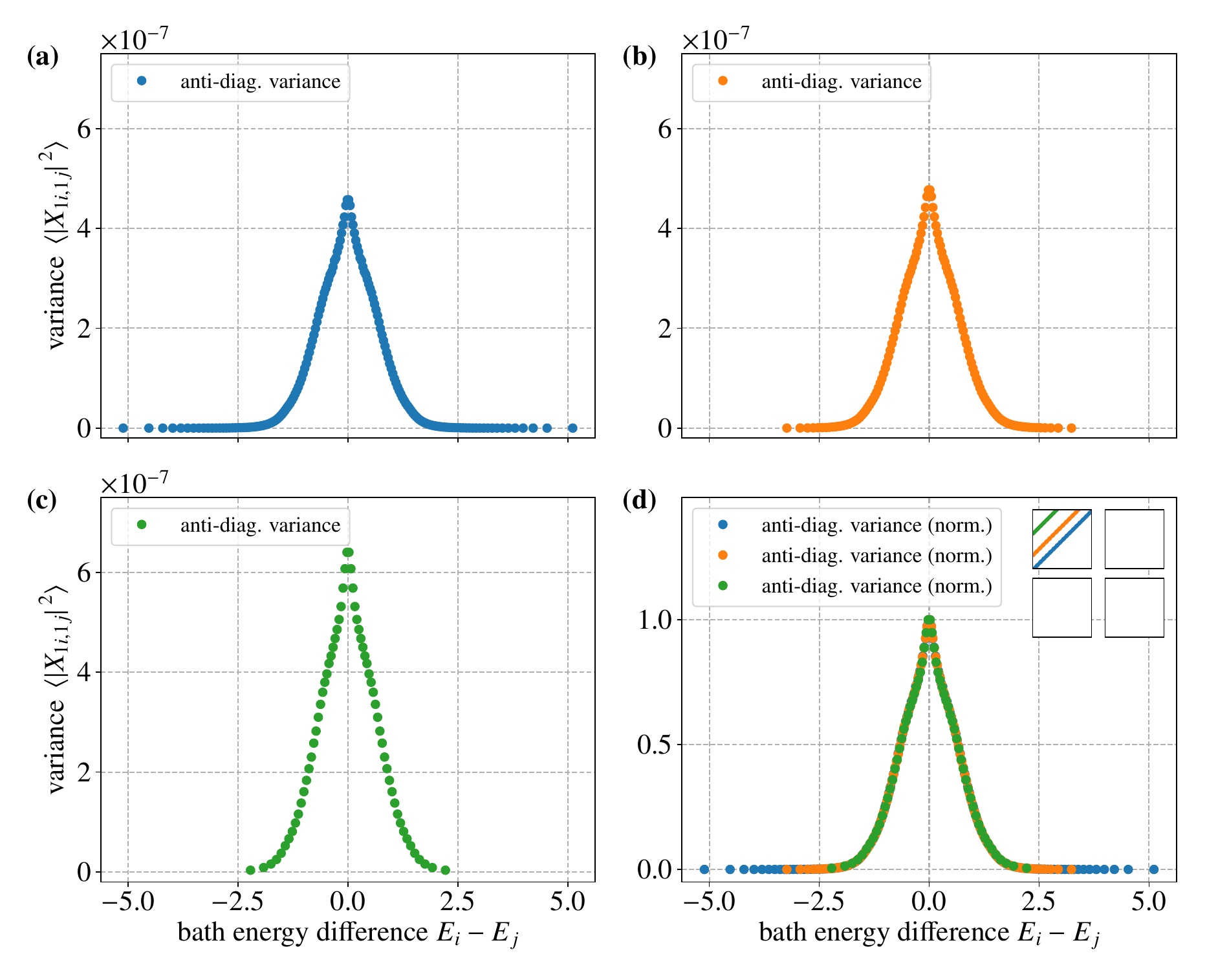}
  \caption{
    \label{fig:12+1sites_anti_dia_comparison}
    Off-diagonal variance $\s_{\mu i,\nu j}^2$ of $\hX$ in a 13 site lattice with $t = \num{2.5e-3}$ for the eigenstate scattering $(\mu,\nu)=(1, 1)$. We extract the variance along three different cross-sections in the anti-diagonal direction in the matrix with an offset of \textbf{(a)} $0\%$, \textbf{(b)} $24\%$ and \textbf{(c)} $61\%$ from the blue anti-diagonal line, as shown in the inset of \textbf{(d)}.
    \textbf{(d)} Variance along the cross-sections in \textbf{(a-c)}, normalized such that the maximum value of each data set is equal to one. The numerical data confirm that the variance along these directions is identical except for an energy dependent prefactor, and hence only dependent on the energy difference $E_i-E_j$.
  }
\end{figure*}

\begin{figure*}[!t]
  \centering
  \includegraphics[width=0.9\linewidth]
  {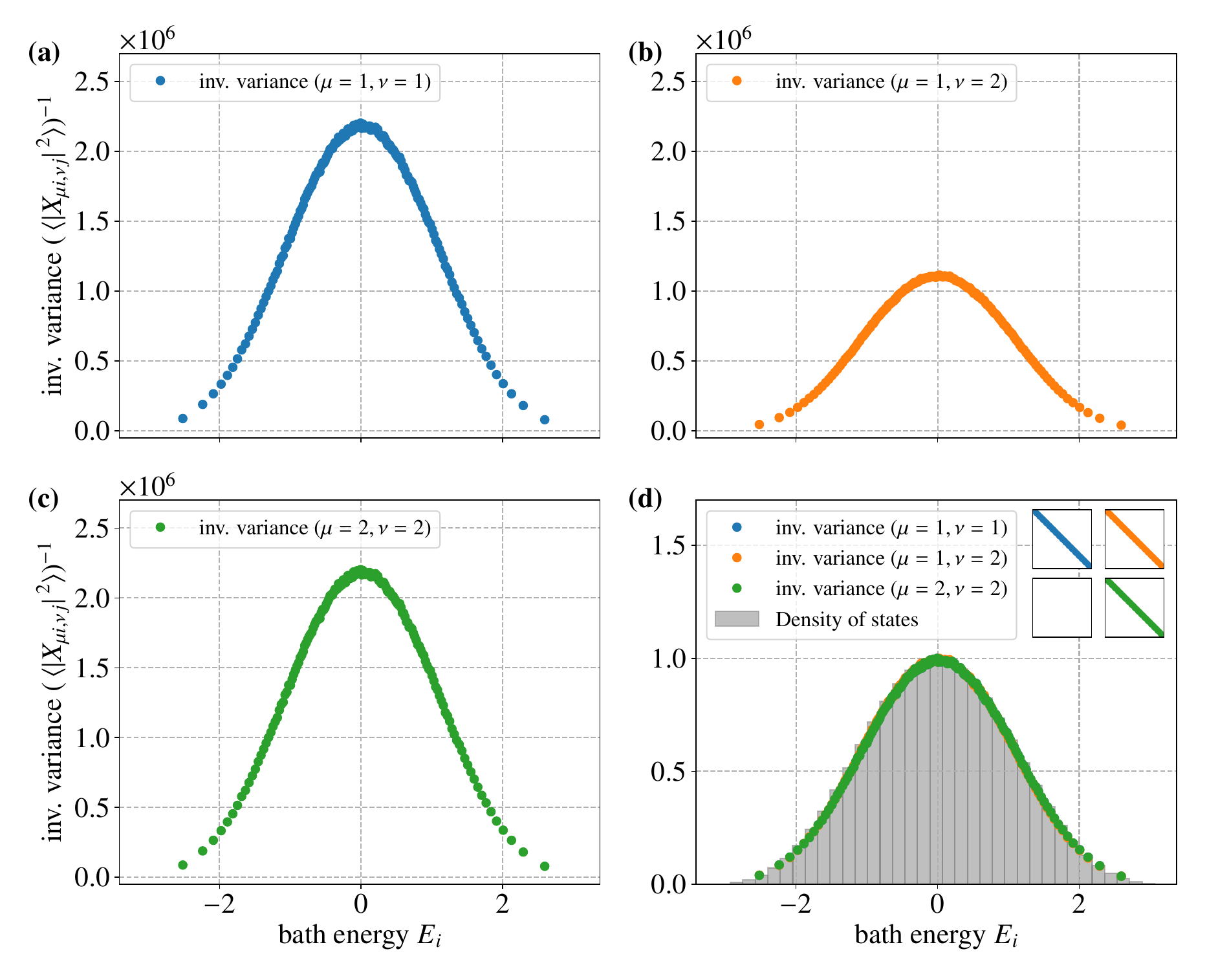}
  \caption{
    \label{fig:12+1sites_dia}
    Off-diagonal variances $\s_{\mu i,\nu j}^2$ of $\hX$ in a 13 site lattice with $t = \num{2.5e-3}$ for the eigenstate scattering $(\mu,\nu)=$ \textbf{(a)} $(1, 1)$, \textbf{(b)} $(1, 2)$ and \textbf{(c)} $(2, 2)$.
    We plot the off-diagonal variance along the diagonal direction of the matrix from the same data as shown in Fig.~\ref{fig:12+1sites_heatmap_energy_log} as a function of the mean bath energy $E_i$, which is equal to $E_j$. \textbf{(d)} Plot of \textbf{(a-c)} together with the eigenvalue density $\rho_\B(E)$ of the bath, normalized such that the maximum position is one for each curve.
    The variance in all four blocks is identical except for a global prefactor. This confirms the factorization 
    \eqref{eq:sigma} as well as the energy dependence of the prefactor in \eqref{eq:tau_ij}.
  }
\end{figure*}

\begin{figure*}[!t]
  \centering
  \includegraphics[width=0.97\linewidth]
  {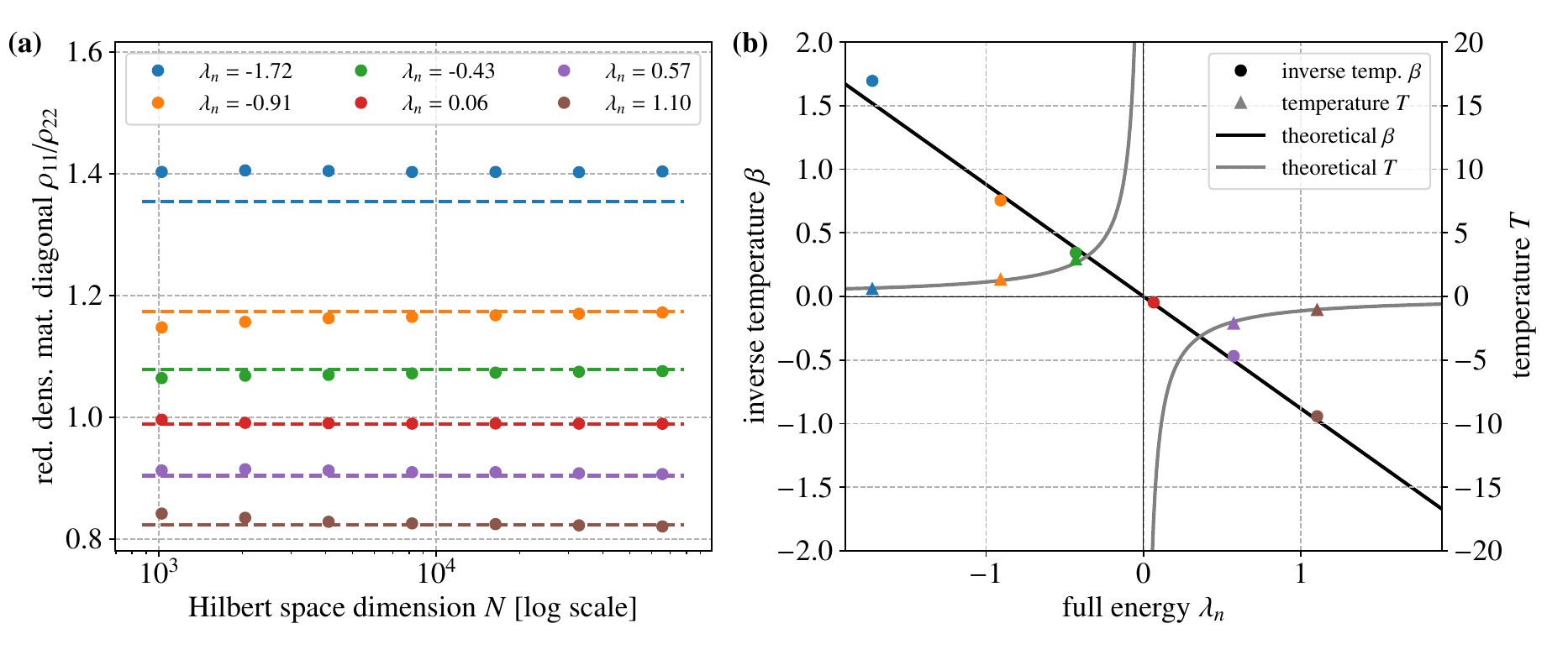}
  \caption{
    \label{fig:red_dens_mat_diagonal}
    \textbf{(a)}
    Ratio of the diagonal values $\rho_{11}$ and $\rho_{22}$ of the reduced density matrix in a two-state subsystem S plotted as a function of the total Hilbert space dimension $N$ (scaled logarithmically) and for different energies $\l_n$. An average is taken over 100 samples of $\hH_\B$ and two samples of $\hat{X}$ for each $\hH_\B$. The dashed line represents the theoretical expectation obtained from the inverse temperature $\beta$, as computed from the density of states with $\D_\tot=\num{1.065}$.  We use the data of Fig.~\ref{fig3}\figletter{c}.
    \textbf{(b)}
    Inverse temperature $\beta$ and temperature $T$ as computed from the data in (a).  The color coding of the data points matches that of {(a)}. The theoretical expectation represented as a solid line computed with $\b= -\l_n/\D_\tot^2$ from the density of states with $\D_\tot=1.065$ shows sufficient agreement with the data.
  }
\end{figure*}

\begin{figure*}[!t]
  \centering
  \includegraphics[width=0.94\linewidth]
  {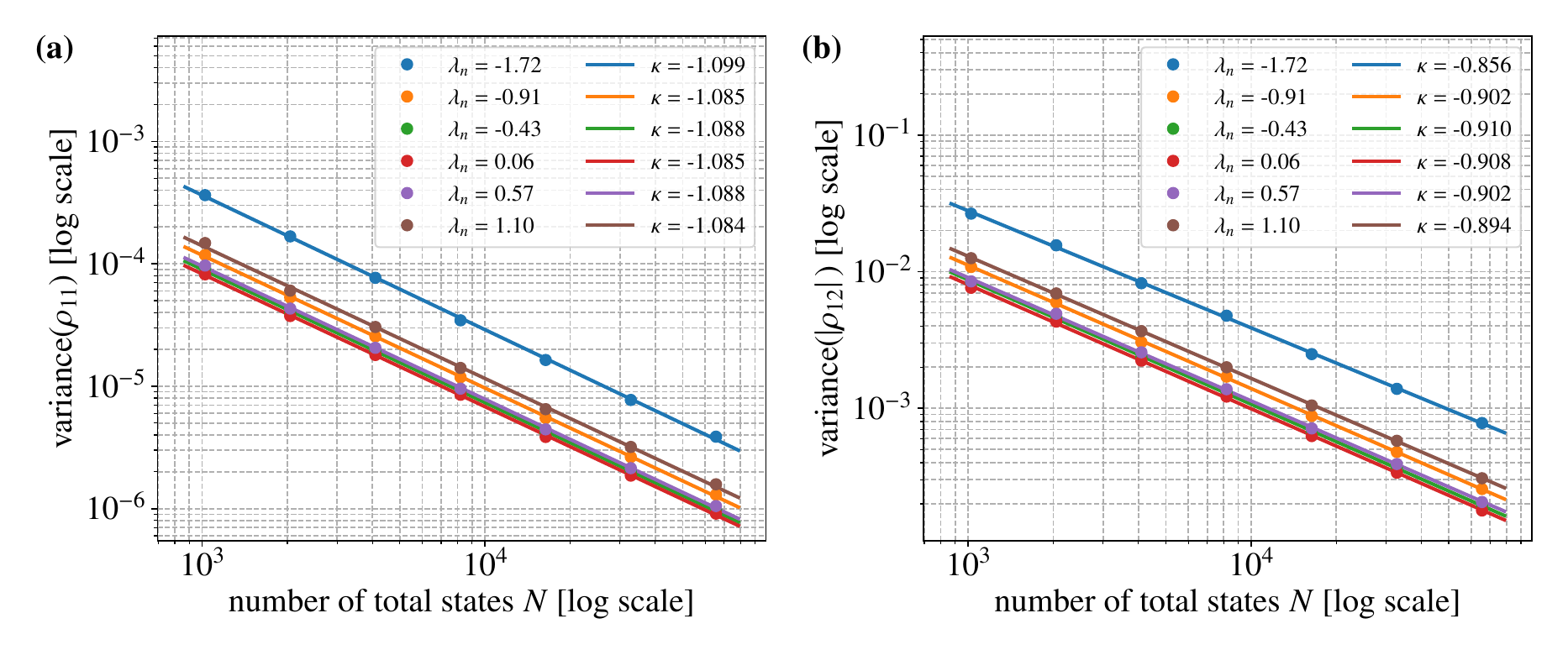}
  \caption{
    \label{fig:red_dens_mat_variance}
    Variance of the entries $\rho_{11}$ and $\abs{\rho_{12}}$ of the reduced density matrix for a two-state subsystem S plotted as a function of the total Hilbert space dimension $N$ in a double-logarithmic plot.
    The variance is computed for the data of Figs.~\ref{fig3}\figletter{c}, \ref{fig3}\figletter{d}, and \ref{fig:red_dens_mat_diagonal} by taking into account the matrix elements for states $\ket{\psi_n}$ in narrow intervals around the desired energy $\l_n$, for all 100 samples of $\hH_\B$, and 200 samples of $\hat{X}$. Fits with exponent $\kappa$ confirm a power law behavior with an exponent close to $\num{-1}$, \ie a scaling $\propto 1/N$.
  }
\end{figure*}

\begin{figure*}[!t]
  \centering
  \includegraphics[width=0.6\linewidth]
  {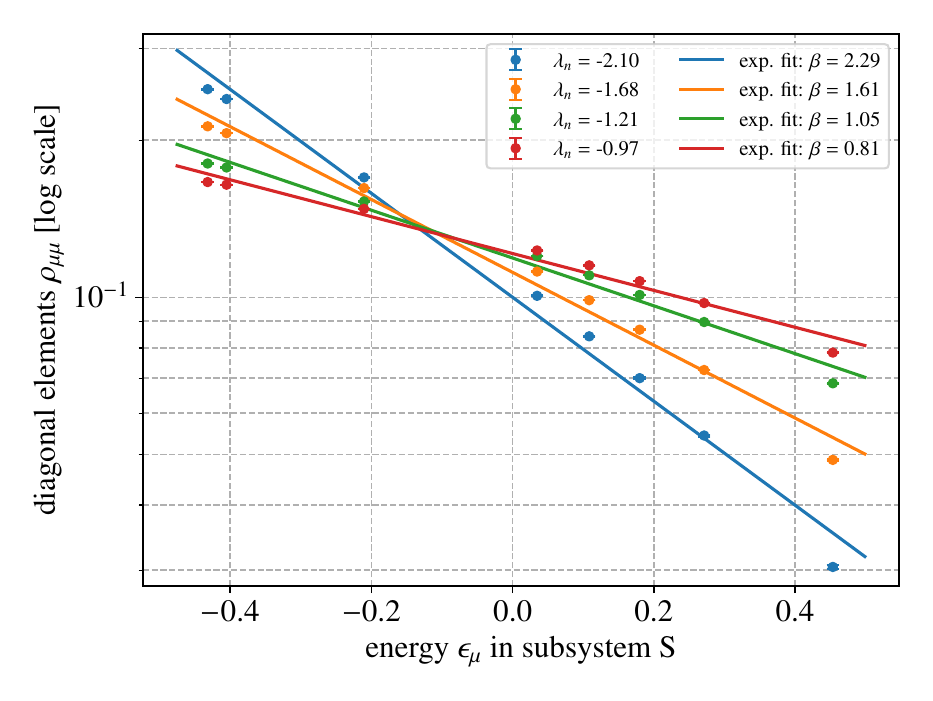}
  \caption{
    \label{fig:red_dens_mat_distribution}
    Diagonal entries $ \rho_{\mu\mu}$ of the reduced density matrix in a system with three sites in S and 13 sites in B on a logarithmic scale. Exponential fits, from which the inverse temperature $\beta$ is extracted, are in good agreement with the data, confirming the Boltzmann distribution of the diagonal elements of the reduced density matrix.
  }
\end{figure*}

Consider a subsystem S coupled weakly to the bath B, and place the entire system in a thermal state with temperature $\b$,
\begin{align}
  \label{eq:A_rho_thermal}
  \hat\rho(\b)=\frac{1}{Z(\b)}\sum_n\,\e^{-\b\l_n}\ket{\psi_n}\bra{\psi_n}.
\end{align}
Now apply the ETH to each eigenstate in the sum,
\begin{align}
  \label{eq:A_ETH}
  \Tr_\B\ket{\psi_n}\bra{\psi_n}=\frac{1}{Z^\S(\b_n)}
  \sum_\mu\,\e^{-\b_n\eps_\mu}\ket{\phi^\S_\mu}\bra{\phi^\S_\mu},
\end{align}
where $Z^\S(\b_n)=\Tr_\S\bigof(\e^{-\b_n\hat H^S})$ is the partition function of the small subsystem S and
$\b_n=-\a\l_n$.  This yields
\begin{align}
  \label{eq:A_TrB_rho_thermal}
  \Tr_\B\of(\hat\rho(\b))
  =&\sum_\mu\,\frac{1}{Z(\b)}\nonumber \\
  \cdot &\sum_n\,\e^{-\b\l_n}
   \frac{1}{Z^\S(\b_n)}
  \e^{-\b_n\eps_\mu}\ket{\phi^\S_\mu}\bra{\phi^\S_\mu},
\end{align}
for the reduced density matrix of the small system.  Consistency requires this to be equivalent to a canonical ensemble with temperature $\b$, \ie
\begin{align}
  \label{eq:A_consist}
  \frac{1}{Z(\b)}\sum_n\,\e^{-\b\l_n}\frac{1}{Z^\S(\b_n)}\,\e^{-\b_n\eps_\mu}
  \overset{!}{=}\frac{1}{Z^\S(\b)}\e^{-\b\eps_\mu}.
\end{align}
Using \eqref{eq:rho_l}, \eqref{eq:bn=-a_ln}, and \eqref{eq:Zb}, we rewrite the l.h.s.\ of \eqref{eq:A_consist} as
\begin{align}
  \label{eq:A_consist_final}
  \sqrt{\frac{\a}{2\pi}}&\intd{\diff\l}
  \e^{-\frac{1}{2}\a\Of{\l+\frac{\b}{\a}}^2}
  \frac{\e^{+\a\l\eps_\mu}}{Z^\S(-\a\l)} 
  \nonumber \\
  &=\frac{1}{\sqrt{2\pi}}\intd{\diff x} \e^{-\frac{1}{2}x^2}
  \frac{\e^{\of{\sqrt{\a}\,x-\b}\eps_\mu}}{Z^\S(-\sqrt{\a}\,x+\b)},
\end{align}
where we have substituted $x\equiv \sqrt{\a}\Of{\l+\frac{\b}{\a}}$.
As we take the limit of a large bath, we recover the r.h.s.\ of \eqref{eq:A_consist} as $\a$ decreases linearly with
the volume $V$ of the bath, $\a\propto \frac{1}{V}$.


\section{Numerical data for the overlaps}
\label{app:lorentzian}

The numerical data for the overlaps $\abs{\!\braket{\phi_{\mu i}|\psi_n}\!}^2$ as shown in Fig.~\ref{fig1}\figletter{b} validate the analytical result, in that they follow a Cauchy-Lorentz distribution close to their peak position. In Fig.~\ref{fig:Lorentzians_log}, we plot the same overlap data for the 18 site system with $t=\num{6.25e-4}$ at energy $\l_n = \num{-0.95789 \pm 0.00025}$ on a logarithmic scale.
Close to the peak positions, the Lorentzian fits in Fig.~\ref{fig:Lorentzians_log} match the data well.  Small deviation arise at overlap values three orders of magnitude smaller than the maximum overlap value. We also notice a slight asymmetry of the overlaps with respect to their maximum position.
To account for this, we defined the center position of the overlap curve, such that the integral in Eq.~\eqref{eq:integral_unity} is equal to the sum rule (completeness of the eigenbases) in Eq.~\eqref{eq:sum_n_chi}.
Since the observed asymmetry is very small, the hereby defined center position agrees with the analytically calculated result in~\eqref{eq:intro_eta_mu} for the shift $\eta_\mu$ in leading order of $t$. Corrections to this formula might apply for larger $t$ with possibly increased asymmetry.

To find the averaged overlap curve in Fig.~\ref{fig1}\figletter{b} and~\ref{fig:Lorentzians_log}, an average is taken over $N_\text{av} = \num{3200}$ full eigenstates $\ket{\psi_n}$ in a small energy interval around $\l_n = \num{-0.95789 \pm 0.00025}$. For this, we sort the overlap values for all states in terms of equally spaced boxes and average over all data in each box individually.
In Fig.~\ref{fig:Lorentzians_statistics} we analyse the statistics of one of the boxes further.
We choose the data point closest to the half-width position of the Lorentzian at $E_i - (\l_n - \eps_1 +\eta_1 ) \approx -\g_1$ (marked in red in Fig.~\ref{fig:Lorentzians_statistics}\figletter{a}) and plot a histogram of the absolute squares of the overlap values in the corresponding box on a logarithmic scale in Fig.~\ref{fig:Lorentzians_statistics}\figletter{b}.
An exponential function with decay constant $\k= \num{591.06}$ fits the histogram.
Thus, the absolute squares of the overlap values at each data point follow an exponential distribution.  This indicates Gaussian distributions of equal variance $\s^2= -1/(2\kappa) =  \num{8.46e-4} $ for the real and imaginary part of the overlap $\braket{\phi_{\mu i}|\psi_n}$.
Since the overlap data points scale inversely as $1/N$ with the Hilbert space dimension $N$ as a result of the completeness of the basis,
the variance $\s^2$ of these data points must scale as $1/N^2$. One can verify both of these statements numerically.

From the Lorentzian fits to the absolute squares of the overlap data, we can extract the half-width $\g_\mu$ and the shift $\eta_\mu$. The results for a 16 site lattice in an average over 400 eigenstates $\ket{\psi_n}$ in energy intervals close to the values of $\l_n$, 100 samples of $\hH_\B$ and two samples of $\hX$ for each $\hH_\B$ are shown in Fig.~\ref{fig3}\figletter{a,b} of the main text in terms of their sums and differences $\g_1 \pm \g_2$ and $\eta_1 \pm \eta_2$.
In Fig.~\ref{fig:Lorentzian_params_vs_E}, we plot the same data for the widths $\g_1$ and $\g_2$ as well as the shifts $\eta_1$ and $\eta_2$ again as a function of the energy $\l_n$. The numerical data are qualitatively described by the analytical results shown as solid lines with shaded regions to indicate the analytical error originating from the uncertainty in the input parameters.
Both $\g_\mu$ and $\eta_\mu$ scale linearly with the interaction strength $t$ (matrix variance of the interaction matrix $\hX$) for all $\mu$ and small $t$.
This is illustrated in Fig.~\ref{fig:Lorentzian_params_vs_t}, where we plot the same data as in Fig.~\ref{fig:Lorentzian_params_vs_E}, but as a function of $t$ instead of $\l_n$.  The linear fits through the origin match the data.
This further confirms that $\g_\mu \rightarrow 0$ and $\eta_\mu \rightarrow 0$ with $t\rightarrow 0$, as expected.
In Fig.~\ref{fig:Lorentzian_params_vs_t}\figletter{c} and \figletter{d} we show that the shifts $\eta_\mu$ occur in opposite directions for $\mu = 1$ and $\mu = 2$, due to the repulsion effect of the interaction $\hX$ described in the main text.
The only exception is the data point at $\l_n = -2.27$, where the thermal shift proportional to the inverse temperature $\beta \propto \l_n$ as an equal contribution for all $\mu$ surpasses the repulsion shift, which causes both Lorentzians to be shifted in the same direction.


\section{Numerical characterization of the interaction matrix $\hX$}
\label{app:xmatrix}



One of the key assumptions in our calculation of the overlaps
$\E\bigof[\abs{\!\braket{\phi_{\mu i}|\psi_n}\!}^2]$ in Sect.~\ref{sec:analytics} is our model for $\s_{\mu i, \nu j}^2$.
It describes the variances of the matrix elements of the perturbation $\hat{X}$ in the unperturbed eigenbasis of S and B and factorizes in a contribution $\s_{\mu \nu}^2$ for the subsystem S and a contribution $\tau_{ij}^2$ for the bath, as formulated in \eqref{eq:sigma}.
In this section, we analyze the matrix elements $X_{\mu i, \nu j} \equiv \bra{\phi_{\mu i}} \hX \ket{\phi_{\nu j}}$ numerically.


Consider the Lorentzian peaks shown in Fig.~\ref{fig1}\figletter{b} and described in~\appendixref{app:lorentzian}. The overlap data are obtained in an 18 site system with $t=\num{6.25e-4}$ for 3200 states $\ket{\psi_n}$ in a narrow energy interval around $\l_n = \num{-0.95789 \pm 0.00025}$.
In Fig.~\ref{fig:17+1sites_X_matrix_Re}, we plot the distribution of the real part of the off-diagonal matrix elements $X_{\mu i, \nu j}$ obtained within boxes of equal energy width $\delta E_i = 0.1$ around the Lorentzian peaks.
For each scattering $(\mu,\nu)$ in S, we choose the center of the boxes at the bath energy positions $(E_i,E_j) = (\l_n-\eps_\mu, \l_n-\eps_\nu)$.
The off-diagonal elements of the boxes are collected in normalized histograms. In all three histograms of Fig.~\ref{fig:17+1sites_X_matrix_Re}, we observe that their real part follows a Gaussian distribution (\cf the Gaussian fits in orange). The variances of the Gaussian fits are all different, due to their different center positions within the bath energies.  This reflects the variance structure of $\hX$ in the unperturbed eigenbasis as shown in Fig.~\ref{fig2}\figletter{a} for $\mu,\nu =1$ and calculated  in~\eqref{eq:tau_ij}.
Due to Hermiticity, the variance for $(\mu,\nu) = (1,2)$ is equal to that for $(\mu,\nu) = (2,1)$.

The histograms for the imaginary part of $X_{\mu i, \nu j}$ shown in Fig.~\ref{fig:17+1sites_X_matrix_Im} are identical to those of the real part in Fig.~\ref{fig:17+1sites_X_matrix_Re}.
This supports our assumption~\eqref{eq:9}, \ie that the real and imaginary part of the matrix elements can be modelled as independent random numbers drawn from Gaussian distributions. 
Since $\hX$ is a local operator, which acts only in a small region R of the lattice, we attribute this behavior to the properties of the eigenbasis of B rather than to the interaction Hamiltonian.
Since the real and imaginary parts of each element $X_{\mu i, \nu j}$ are both described by Gaussian distributions with standard deviation $\s$, the absolute squares $\abs{X_{\mu i, \nu j}}^2$ follow an exponential distribution. This is shown in Fig.~\ref{fig:17+1sites_X_matrix_Abs2} on a logarithmic scale and confirmed by exponential fits.
The decay constants $\k$ of the fits given by $\k = -1/(2 \s^2)$ agrees with the values for $\s$ as determined in Fig.~\ref{fig:17+1sites_X_matrix_Re} and Fig.~\ref{fig:17+1sites_X_matrix_Im} for each plot, respectively.

In Fig.~\ref{fig:17+1sites_X_matrix_Dia}, we evaluate the histograms for the diagonal elements $X_{\mu i, \mu i}$ in the two boxes with $\mu = \nu$.
They also follow Gaussian distributions with variances approximately twice as large as the variances of the real and imaginary part in the off-diagonal elements of the same box (\cf Fig.~\ref{fig:17+1sites_X_matrix_Re}\figletter{a,c} and Fig.~\ref{fig:17+1sites_X_matrix_Im}\figletter{a,c}). Combining the results, the squared boxes around the diagonal for $\mu = \nu$ resemble (different) Gaussian unitary ensembles within each of the given small energy intervals.
This validates the form of the ansatz~\eqref{eq:9}, where the individual variances for the real and imaginary part of $X_{\mu i, \nu j}$ are given by $\s_{\mu i, \nu j}^2/2$ for $\mu \neq \nu$ and $i\neq j$, while the variance of $X_{\mu i, \mu i}$ is $\s_{\mu i, \mu i}^2$.
This results in the simple relation $\E[\abs{X_{\mu i, \nu j}}^2] = \s_{\mu i, \nu j}^2$.



The variance structure of the whole matrix $X_{\mu i, \nu j}$ shown in Fig.~\ref{fig:12+1sites_heatmap_log} is resolved in terms of the indices of the bath eigenstates $i,j$, with the four plots corresponding to the four scatterings $(\mu,\nu)$ of the eigenstates in S. For this analysis, we consider a 13 site lattice with $t=\num{2.5e-3}$ and an average over 100 random samples of $\hH_\B$ and two random samples of $\hX$ for each $\hH_\B$.
The heat maps for $(\mu,\nu) = (1,1), (1,2), (2,1)$ and $(2,2)$ in Fig.~\ref{fig:12+1sites_heatmap_log} appear identical except for an overall scaling factor. We observe a ``leaf-shaped'' distribution, with the variance decreasing in an elliptic shape for elements that are both further away from the diagonal and further away from the boundary of the spectrum.

Since $\hat{X}$ acts inside a small region R of B, it can change the energy of bath eigenstates only by the small amount available to the region R, independently of the scattering in S. As a consequence, the scattering amplitude decreases for states associated with a larger energy difference (\ie along the anti-diagonal). In addition, the variance model has to fulfill the normalization property~\eqref{eq:tau_norm}, where the sum over each row or column is normalized to one. Since the eigenvalue density is a Gaussian, the mean energy difference between neighboring states increases towards the boundary, resulting in a smaller number of eigenstates inside the scattering width.
Hence, the variance must increase towards the boundary of the spectrum with a suppressed eigenvalue density and fewer scattering elements available, which explains the pointed ends of the ``leaf''.

When the plot is recast in terms of the bath energies $E_i$ and $E_j$ instead of the indices $i$ and $j$ in Fig.~\ref{fig:12+1sites_heatmap_energy_log}, the band character of the variance model becomes visible. It shows, that the scattering width stays constant throughout the spectrum resulting in concise energetic lines of constant scattering amplitude. This suggests that the variances are a function of the bath energies $E_i$ and $E_j$. Furthermore, we find a ``quasi-periodicity'' of the scattering in the bulk of the spectrum where the variance is, up to a normalization factor, determined by a function of the energy difference $E_i-E_j$.


To discern the functional dependence of the variance model on the energy difference $E_i-E_j$, we extract the variance for a line along the anti-diagonal direction of the matrix for all scattering sectors $(\mu, \nu)$ in Fig.~\ref{fig:12+1sites_anti_dia}.
The figure suggests that the variance $\t_{ij}^2$ in B corresponds roughly to a Gaussian with an additional peak at zero, which we attribute to finite-size effects.
Due to the construction of the Hamiltonian in~\eqref{eq:num_ham}, with twice as many off-diagonal scattering terms ($\hat{\s}^1$ or $\hat{\s}^2$ acting on site 1) of $\hX$ in S as diagonal ones ($\hat{\s}^3$ acting on site 1), the variances for $(\mu, \nu) = (1, 1)$ and $(\mu, \nu) = (2, 2)$ are smaller than the variances in $(\mu, \nu) = (1, 2)$.
Specifically, we expect $\s_{1,1}^2 = \s_{2,2}^2 = 1/3$ and $\s_{1,2}^2 = \s_{2,1}^2 = 2/3$.
This is numerically confirmed in Fig.~\ref{fig:12+1sites_anti_dia}\figletter{d}, where $2\times \s_{1 i,1 j}^2$ and $2\times \s_{2 i,2 j}^2$ agree with $\s_{1 i,2 j}^2$.
The results additionally verify that the scattering of $\hX$ in S and B is independent, resulting in a factorization of the variance through $\s_{\mu i,\nu j}^2 =\s_{\mu\nu }^2\hspace{1pt} \t_{ij}^2$ as described in~\eqref{eq:sigma}.

Fig.~\ref{fig:12+1sites_anti_dia_comparison} shows the variance of $X_{1 i, 1 j}$ with $\mu,\nu=1$ in S for three different cross-sections parallel to the anti-diagonal line in the matrix. We observe that the variance increases by a scaling prefactor as we move towards the boundary of the spectrum, while its overall shape remains the same.  This is best seen in Fig.~\ref{fig:12+1sites_anti_dia_comparison}\figletter{d}, where the variance along the three cross-sections is normalized such that their maximum value is equal to one.
The results confirm the ``quasi-periodicity'' in $X_{\mu i, \nu j}$, with the variance $\t_{ij}^2$ in B depending solely on the bath energy difference $E_i-E_j$ and an energy-dependent overall scaling factor, as reflected in~\eqref{eq:tau_ij}.

To determine the scaling factor in $\t_{ij}^2$, we plot the inverse of the off-diagonal variance of the matrix $X_{\mu i, \nu j}$ along its diagonal direction in Fig.~\ref{fig:12+1sites_dia} for different values of $\mu,\nu$. In this plot, we see again that $\t_{ij}^2$ is twice as large for $\mu\neq \nu$ as compared to $\mu=\nu$.
When normalized such that the maximum value is one in Fig.~\ref{fig:12+1sites_dia}\figletter{d}, the data agrees with the density of states of the bath $\rho_\B(E)$, being normalized in the same way. Thus, we conclude that the scaling factor is inversely proportional to the eigenvalue density. To symmetrize the result in the indices $i$ and $j$, we incorporate the square root of the product of the eigenvalue density at $E_i$ and $E_j$ in the variance model (\cf Eq.~\eqref{eq:tau_ij}). The numerical results validate our Ansatz for $\s_{\mu i, \nu j}^2$.

Finally, note that the figures~\ref{fig:12+1sites_anti_dia} or~\ref{fig:12+1sites_anti_dia_comparison} are also used to obtain the value of the scattering width $\Delta_0$, which is an input parameter in the analytical expressions for $\gamma_\mu$ and $\eta_\mu$ given in~\eqref{eq:intro_gamma_mu} and~\eqref{eq:intro_eta_mu}. For this, we perform a Gaussian fit as shown in Fig.~\ref{fig2}\figletter{b}.


\section{Variances $\tau_{ij}^2$ of the elements in $\hX$}
\label{app:tau_variances}


We now derive \eqref{eq:tau_ij} under the following assumptions.  First, there is no correlation between the scattering elements in the small matrix $\hX^{\S\RR}$
which acts only on the part of the Hilbert space describing S and R,
and the energy of the states in R.  Second, the eigenvalue density of the region R is given by a Gaussian,
\begin{align}
  \label{eq:rho_R}
  \rho_\RR\of(\eps)=\frac{1}{\sqrt{2\pi} \D_\RR}
  \exp\Of{-\frac{\eps^2}{2\D_\RR^2}}.
\end{align}
Third, the coupling between R and the remainder of the bath $\B\setminus\RR$ is infinitesimal.

For this calculation, we use the notation of the main text but with the system S now representing the region R, the bath B the region $\B\setminus\RR$, and the entire system $\S\cup\B$ the bath B. In this language, we are interested in the matrix elements
\begin{align}
  \label{eq:Y_nm}
  Y_{nm}\equiv\bra{\psi_n}\hY^\S\otimes\1^\B\ket{\psi_m}
\end{align}
of an perturbation $\hY^\S$ with unit matrix variance acting only on S, or more precisely, in the variances of these elements,
\begin{align}
  \label{eq:tau_nm_sq}
  \tau_{nm}^2=\E\Of[\bra{\psi_n}\hY^\S\otimes\1^\B\ket{\psi_m}
                   \bra{\psi_m}\hY^\S\otimes\1^\B\ket{\psi_n}].
\end{align}
Since we assume that the matrix elements 
$Y_{\mu\nu}^\S=\bra{\phi^\S_\mu}\hY^\S\ket{\phi^\S_\nu}$
are uncorrelated, \linebreak
$\E\of[Y_{\mu\nu}^\S Y_{\nu'\mu'}^\S]=\d_{\mu\mu'}\d_{\nu\nu'}\varsigma_{\mu\nu}^2$ with normalization
$\sum_{\mu\nu}\varsigma_{\mu\nu}^2=N_\S$, we may rewrite \eqref{eq:tau_nm_sq} as
\begin{align}
  \label{eq:tau_nm_sq_2}
  \tau_{nm}^2= &\sum_{\mu\nu}\varsigma_{\mu\nu}^2 \nonumber \\
  \cdot &\sum_{ij} \E\Of[
  \braket{\psi_n|\phi_{\mu i}}\braket{\phi_{\nu i}|\psi_m}
  \braket{\psi_m|\phi_{\nu j}}\braket{\phi_{\mu j}|\psi_n}]
\end{align}
Since the phases of the overlaps are only correlated for $i=j$,
we obtain
\begin{align}
  \label{eq:tau_nm_sq_3}
  \tau_{nm}^2=\sum_{\mu\nu}\varsigma_{\mu\nu}^2 \sum_i \chi_{\mu i,n}\chi_{\nu i,m}.
\end{align}
For the last sum, we refer to \eqref{eq:S_mu_nu_squared}.
Since we assume an infinitesimal coupling $t$, we take $\eta_\mu,\g_\mu\to 0$, such that the Cauchy distribution turns into
$\d\of{\o+\eps_\mu-\eps_\nu}$ with $\o=\l_m-\l_n$.
%
With \eqref{eq:Zb} and $b_\RR\equiv\frac{1}{4}\b \D_\RR$,
\begin{align}
  \label{eq:Z^S}
  Z^\S=N_\S\exp\Of{\half\b^2\D_\RR^2}
  =N_\S\e^{8\hspace{.5pt} b_\RR^2}.
\end{align}
Replacing the sums over $\mu$ and $\nu$ by integrals, we obtain
\begin{align}
  \label{eq:tau_nm_sq_4}
  &\tau_{nm}^2=\frac{N_\S\e^{-8\hspace{.5pt} b_\RR^2}}{N\sqrt{\rho\of{\l_n}\rho\of{\l_m}}} \nonumber \\
  \cdot &\intd{\diff\eps\diff\eps'}\rho_\RR(\eps)\rho_\RR(\eps')
  \hspace{1pt}\varsigma^2(\eps,\eps') \hspace{1pt}
  \e^{-\half\b(\eps+\eps')}
  \d\of{\o+\eps-\eps'}.
\end{align}
Due to our assumption that the variances of $\hY^\S$ are not correlated with the energies of the states in R they scatter, we may replace
$\varsigma^2(\eps,\eps')\to\sfrac{1}{N_\S}$ in the integral.  With \eqref{eq:rho_R},
\begin{align}
  \nonumber
  \rho_\RR\of(\eps)\e^{-\half\b\eps}
  =\rho_\RR\Of(\eps+\half\b\D_\RR^2)\e^{2\hspace{.5pt} b_\RR^2}, 
\end{align}
such that
\begin{align}
  \label{eq:tau_nm_sq_5}
  \tau_{nm}^2
  &=\frac{\e^{-4\hspace{.5pt} b_\RR^2}}{N\sqrt{\rho\of{\l_n}\rho\of{\l_m}}}
  \intd{\diff\eps} \rho_\RR\Of(\eps) \rho_\RR\Of(\eps+\o)
    \nonumber\\[5pt]
  &=\frac{\e^{-4\hspace{.5pt} b_\RR^2}}{N\sqrt{\rho\of{\l_n}\rho\of{\l_m}}}
  \frac{1}{\sqrt{\pi}\hspace{1pt}2\D_\RR}\exp\Of{-\frac{\o^2}{4\D_\RR^2}}.
\end{align}
With $\tau_{nm}\to\tau_{ij}$, $N\to N_\B$, $\rho\to \rho_\B$, $\l_n\to E_i$, $\l_m\to E_j$, $\o\to E_j-E_i$, and $\D_\RR\to\half\D_0$, $b_\RR\to\half b_0$,
we recover \eqref{eq:tau_ij}.

If the coupling $t_\RR$ between the region R and the remainder of the bath $\B\setminus\RR$ is not small, we need to replace the $\delta$-function in \eqref{eq:tau_nm_sq_4} by a Lorentzian with half-width $\g_\mu+\g_\nu$, and no analytic solution is possible.  An approximate way to account for the coupling is to devide the matrix variance $t_\RR$ equally between R and $\B\setminus\RR$, \ie to replace $\D_\RR^2\to \D_\RR^2+\half t_\RR $ in \eqref{eq:tau_nm_sq_5}.


\section{Numerical results for the reduced density matrix}
\label{app:red_dens_mat}

With our analytical results on the eigenvector overlaps, we derived the canonical ensemble within the subsystem S by considering the reduced density matrix in S, obtained from the 
density matrix of a single quantum state. In the main text, we have shown that the diagonal entries of the reduced density matrix describing S follow a Boltzmann distribution with shifted energy levels, while its off-diagonal entries scale with $1/\sqrt{N}$, and hence decrease exponentially with the system size of the bath. In this section, we present
numerical evidence and confirmation.

The diagonal elements of the reduced density matrix are shown in Fig.~\ref{fig:red_dens_mat_diagonal} for a system with a total of $\{ 10, 11, 12, 13, 14, 15, 16 \}$ lattice sites and
interaction strengths $t=\{\num{5.92}, \num{5.08}, \num{4.44}, \num{4.10}, \num{3.77}, \num{3.36}, \num{3.13}\} \times 10^{-3}$.
We take an average over $N_\text{av} = \{ 50,$ $100, 200, 400, 800, 1600, 3200 \}$ states $\ket{\psi_n}$ in narrow intervals around the desired energy $\l_n$, as well as an additional average over 100 samples of $\hH_\B$ and two samples of $\hat{X}$ for each $\hH_\B$. The averaged energy intervals around $\l_n$ are held approximately constant independent of the system size.
In Fig.~\ref{fig:red_dens_mat_diagonal}\figletter{a}, we plot the ratio $\rho_{11}/\rho_{22}$ with respect to the Hilbert space dimension $N$, as 
in Fig.~\ref{fig3}\figletter{c} in the main text. The ratio stays constant with system size and agrees with the theoretical prediction of Boltzmann factors, with $\beta$ computed from the eigenvalue density of width $\D_\tot=1.065$, with small deviations at $\l_n=-1.72$ due to the suppressed eigenvalue density at the spectral boundary.

From the diagonal of the reduced density matrix in subsystem S, we can compute the inverse temperature $\beta$ by fitting the corresponding Boltzmann distribution.
The results are shown in Fig.~\ref{fig:red_dens_mat_diagonal}\figletter{b}.
The solid lines in Fig.~\ref{fig:red_dens_mat_diagonal}\figletter{b} represent the theoretical expectation according to $\beta = -\l_n/\D_\tot^2$ and matches the numerical results. This confirms the thermal behavior of subsystem S in the quantum system at an exact eigenstate with the inverse temperate computed as the energy derivative of the entropy $S(E) \propto - \ln(\rho(E))$.

In addition to the average values of the reduced density matrix, we have evaluated their variance as a function of the total Hilbert space size $N$.  The results are shown in Fig.~\ref{fig:red_dens_mat_variance} in a double-logarithmic plot.
The variance of both the diagonal elements $\rho_{11}$ (Fig.~\ref{fig:red_dens_mat_variance}\figletter{a}) and the off-diagonals $\rho_{12}$  (Fig.~\ref{fig:red_dens_mat_variance}\figletter{b}) follow a power law in the Hilbert space dimension $N$, which corresponds to an exponential decrease with the system size $L$.
In contrast to the averages of $\abs{\rho_{12}}$, which fall off as $1/\sqrt{N}$, the variances of $\rho_{11}$ and $\rho_{12}$ decrease approximately as $1/N$, as one can see from the fitted exponents $\kappa \approx 1$.

Finally, we consider a quantum system with three sites in S and 13 sites in B.
For this system, Fig.~\ref{fig:red_dens_mat_distribution} displays the thermal occupations $ \rho_{\mu\mu}$ of the numerically obtained reduced density matrix.
The 8 states in S follow roughly an exponential distribution for all state energies $\l_n$, as demonstrated by the exponential fits. The decay constant
$\beta$ corresponds 
to the inverse temperature. 


\section{Validity of Ansatz \eqref{eq:chi_mu_i_n_ansatz} for strong coupling}
\label{app:validity}

In Sect.~\ref{sec:analytics}, we found that the canonical ensemble for subsystem S, and in particular the diagonal entities for the reduced density matrix \eqref{eq:rho_mu_mu}, does not depend on the details of the solution of the Casati--Girko formalism, but only on the validity of Ansatz \eqref{eq:chi_mu_i_n_ansatz} for the overlap functions.  As we obtained the analytic solution for weak coupling matrices $\hX^t$, we found that
$\tG_{\mu i}^t(\l_n-\i0^+)$, and hence $\chi_\mu(x)$, depends on $a_{\mu i}$ and $\l_n$ only through $x=a_{\mu i}-\l_n-\eta_\mu$ in $v=\frac{1}{\D_1}(x-\eps_{\mu\nu})+b_1$ (see \eqref{eq:Wieltjes} and \eqref{eq:chi_mu_form}), as
asserted
in \eqref{eq:chi_mu_i_n_ansatz}.  In this \appendixname , we will show that this dependence,
and hence the canonical ensemble, 
prevails even if the coupling is strong, and the problem not amenable to analytic solution.

The only assumption we need to make is that the variances of the scattering elements $\t_{ij}^2$ depend, apart form the normalization prefactors related to the eigenvalue densities $\rho_\B(E_i)$ and $\rho_\B(E_j)$, only on the difference $E_i-E_j$, \ie
\begin{align}
  \label{eq:tau_ij_general}
  \t_{ij}^2= \frac{f\of{E_i-E_j}}{N_\B\sqrt{\rho_\B(E_i)\rho_\B(E_j)}},
\end{align}
where the peak function $f(\omega)$ 
has be symmetric and consistent with the normalization \eqref{eq:tau_norm}, which now reads
\begin{align}
  \label{eq:tau_norm_general}
  \sum_{i} \t_{ij}^2
    =\intd{\diff\omega} e^{+\half\b\omega}f(\omega)
    =1.
\end{align}
A very reasonable choice for $f(\omega)$ is the Gaussian of \eqref{eq:tau_ij} we derived in \ref{app:tau_variances}, but no specific choice is required for the argument.

With this assumption, we now show that if we start the Casati--Girko iteration with an Ansatz for the overlap function of the form \eqref{eq:chi_mu_form},
\begin{align}
  \label{eq:chi_mu_i_n_ansatz_general}
  \chi_{\mu i,n}=\frac{\chi_\mu^{\initial}(a_{\mu i}-\l_n-\eta_\mu)}
  {N\sqrt{\rho(a_{\mu i}-\eta_\mu)\rho(\l_n)}},
\end{align}
we will obtain an overlap function $\chi_\mu
(a_{\mu i}-\l_n-\eta_\mu)$ which likewise depends only on the combination $a_{\mu i}-\l_n-\eta_\mu$.  Using \eqref{eq:Im_R}, we find
\begin{align}
  \label{eq:Im_R_general}
  \frac{1}{\pi}\Im\R_{\nu j}^t(\xi-\i 0^+)
  =e^{-\half\b\of{a_{\nu j}-\eta_\nu -\xi}}\,\chi_\nu^{\initial}(a_{\nu j}-\xi-\eta_\nu).
\end{align}
Substitution into \eqref{eq:Im_Gtilde} yields
\begin{align}
  \label{eq:Im_Gtilde_general}
  &\frac{1}{\pi}\Im\tG_{\mu i}^t(\xi-\i 0^+)
  =\sum_\nu \s_{\mu\nu}^2 \sum_j \tau_{ij}^2\,
   \frac{1}{\pi}\Im\R_{\nu j}^t(\xi-\i 0^+) \nonumber\\
  &=\sum_\nu\s_{\mu\nu}^2 \hspace{1pt}\intd{\diff a}
    e^{+\half\b\of{(a-\eps_\nu)-(a_{\mu i}-\eps_\mu)}} \nonumber\\[-4pt]
  &\hspace{12pt}  \cdot f\Of{(a-\eps_\nu)-(a_{\mu i}-\eps_\mu)}
    e^{-\half\b\of{a-\eta_\nu -\xi}}\,
      \chi_\nu^{\initial}(a-\xi-\eta_\nu)
      \nonumber\\[5pt]
  &=\sum_\nu\s_{\mu\nu}^2 \hspace{3pt}
    g_\nu\of{a_{\mu i}-\xi-\eta_\mu-\eps_{\mu\nu}},
\end{align}
where $\eps_{\mu\nu}\equiv(\eps_\mu-\eta_\mu)-(\eps_\nu-\eta_\nu)$ and
\begin{align}
  \label{eq:xi_general}
  g_\nu\of{\zeta}
  \equiv e^{-\half\b\zeta}\intd{\diff\omega}f\of{\omega}\chi_\nu^{\initial}(\omega+\zeta).
\end{align}
Using the Kramers--Kronig relation \eqref{eq:KK}, we obtain
\begin{align}
  \label{eq:Re_Gtilde_general}
  \Re\tG_{\mu i}^t(\l_n)
  =\sum_\nu \s_{\mu\nu}^2\hspace{1pt}
  \dashint\hspace{-2pt}\diff \pu\hspace{3pt}\frac{g_\nu\of{\pu +\pv }}{\pu },
\end{align}
where 
we have defined $\pu\equiv\l_n-\xi$,
$\pv\equiv a_{\mu i}-\l_n-\eta_\mu-\eps_{\mu\nu}$.  Combining \eqref{eq:Im_Gtilde_general} and \eqref{eq:Re_Gtilde_general}, we obtain the Wieltjes transformation
\begin{align}
  \label{eq:Wieltjes_general}
  \tG_{\mu i}^t(\l_n-\i0^+)
  =\sum_\nu\s_{\mu\nu}^2\biggl(\,
  \dashint\hspace{-2pt}\diff \pu\hspace{3pt}\frac{g_\nu\of{\pu +\pv }}{\pu }
  + \i\pi\hspace{1pt} g_\nu\of{\pv }
  \biggr).
\end{align}
As we substitute \eqref{eq:Wieltjes_general} into \eqref{eq:chi_mu_form}, we see that $\chi_\mu(x)$ depends on $a_{\mu i}$ and $\l_n$ only through the combination $x\equiv a_{\mu i}-\l_n-\eta_\mu$ (using $\pv =x-\eps_{\mu\nu}$).

\vspace{0.7cm}


\section{Lorentzian half-width $\g(\eps)$ and shift $\eta(\eps)$ for a large subsystem S}
\label{app:large_system_S}

We now assume that the system S is still small compared to B, but large enough that its eigenvalue density is given by a Gaussian with variance $\D_\S^2$, and that $\hX$ scatters only degrees of freedom in a small region Q of S.  The variances $\s_{\mu\nu}^2$ will take a form similar to \eqref{eq:tau_ij} for $\tau_{ij}^2$.  The difference is that we need $\s_{\mu\nu}^2$ for all energies, and hence all temperatures, of the finite system S, while we could assume an infinite bath B, and hence a constant temperature $\b$, when we evaluated $\tau_{ij}^2$ in \appendixref{app:tau_variances}.

To calculate $\s_{\mu\nu}^2$, we hence redo the calculation, but now assume Gaussian eigenvalue densities \eqref{eq:rho_R} not only  with variance $\D_\Q^2$ for the region Q, but also for B and the entire system, with respective variances $\D_\B^2$ and 
$\D_{\tot}^2$.
We make the same three assumptions as in \appendixref{app:tau_variances}, and again use the notation of the main text, but with the system S now representing the region Q, the bath B the region $\S\setminus\Q$, and the entire system $\S\cup\B$ the system S.  In this language, we find again \eqref{eq:tau_nm_sq_3},
\begin{align}
  \label{eq:gauss_tau_nm_sq_1}
  \tau_{nm}^2=\sum_{\mu\nu}\varsigma_{\mu\nu}^2 \sum_i \chi_{\mu i,n}\chi_{\nu i,m},
\end{align}
but 
this time, we have to evaluate the sum over $i$ using a Gaussian eigenvalue density.
With $\chi_{\mu i,n}={\d\of{a_{\mu i}-\l_n}}/{N\rho\of{\l_n}}$, we obtain
\begin{align}
  \label{eq:gauss_sum_i}
  \sum_i \chi_{\mu i,n}\chi_{\nu i,m}=
  \frac{1}{N_\S N}\,\frac{\rho_\B\of{\l_n-\eps_\mu}}{\rho\of{\l_n}\rho\of{\l_m}}
  \,\d\of{\o+\eps_\mu-\eps_\nu},
\end{align}
where $\o=\l_m-\l_n$.  This yields
\begin{widetext}
\begin{align}
  \label{eq:gauss_tau_nm_sq_2}
   \tau_{nm}^2&=\frac{N_\S}{N{\rho\of{\l_n}\rho\of{\l_m}}}
                \intd{\diff\eps\diff\eps'}\rho_\Q(\eps)\rho_\Q(\eps')
                \,\varsigma^2(\eps,\eps')\,\rho_\B\of{\l_n-\eps} \d\of{\o+\eps-\eps'}
  \nonumber\\
              &=\frac{1}{N{\rho\of{\l_n}\rho\of{\l_m}}}
                \intd{\diff\eps}\rho_\Q(\eps)\rho_\Q(\eps+\o)\,\rho_\B\of{\l_n-\eps}.
\end{align}
\end{widetext}
Substituting Gaussians with variances $\D_\Q^2$, $\D_\B^2$, and $\D_{\tot}^2=\D_\B^2+\D_\Q^2$ for the eigenvalue densities $\rho_\Q$, $\rho_\B$, and $\rho$, we obtain
\begin{align}
  \label{eq:gauss_tau_nm_sq_3}
   \tau_{nm}^2&=\frac{1}{N\rho\of{\l_m}} \frac{1}{\sqrt{\pi}\hspace{1pt}\D_0'}
        \exp\Of[-\frac{1}{{\D_0'}^2}\left(\o+\l_n\frac{\D_\Q^2}{\D_{\tot}^2} \right)^2], 
\end{align}
where
\begin{align}
  \label{eq:gauss_D0_prime}
   \D_0'=2\D_\Q\,\sqrt{1-\frac{\D_\Q^2}{2\D_{\tot}^2}}.
\end{align}
Note that since $\l=-\b\D_{\tot}^2$ (see \eqref{eq:bn=-a_ln}), in the large bath limit $\D_{\tot}\gg \D_\Q$, \ref{eq:gauss_tau_nm_sq_3} becomes
\begin{align}
  \label{eq:gauss_tau_nm_sq_4}
   \tau_{nm}^2&=\frac{1}{N\rho\of{\l_m}} \frac{1}{\sqrt{\pi}\hspace{1pt}\D_0}
        \exp\Of[-\frac{1}{{\D_0}^2}\left( \o-\frac{\b}{4} {\D_0}^2 \right)^2],
\end{align}
which is equivalent to \eqref{eq:tau_nm_sq_5} with $\D_0=2\D_\RR$.

With $\tau_{nm}\to\s_{\mu\nu}$, $N\to N_\S$, $\rho\to \rho_\S$, $\l_n\to\eps_\mu$, $\l_m\to\eps_\nu$, $\o\to \eps_\nu-\eps_\mu$, and $\D_{\tot}\to\D_\S$, we obtain
\begin{align}
  \label{eq:gauss_sigma_sq_1}
  \s_{\mu\nu}^2=\frac{1}{N_\S\rho_\S\of{\eps_\nu}}
  \frac{1}{\spi\hspace{1pt}\D_0'}
  \exp\Of[-\frac{1}{{\D_0'}^2}\bigl(\eps_\nu-\eps_\mu (1-\q)\bigr)^2],
  \nonumber \\
\end{align}
where $\q\equiv\sfrac{\D_\Q^2}{\D_\S^2}$
denotes the ratio between the variances of the region Q and the system S, and
\begin{align}
  \label{eq:gauss_D0_prime2}
   \D_0'=2\D_\Q\,\sqrt{1-\half \q\,}.
\end{align}

To calculate $\g\of(\eps)$ and $\eta\of(\eps)$ to first order in $t$, we substitute \eqref{eq:gauss_sigma_sq_1} into \eqref{eq:Im_Gtilde} with $\D_\p=0$,
\begin{align}
  \label{eq:gauss_Im_Gtilde}
  &\frac{1}{\pi}\Im\tG_{\mu i}^t(\xi-\i 0^+) \nonumber \\
  & =\frac{1}{\pi\D_0 \hspace{1pt}\D_0'}
  \intd{\diff\eps}
  \exp\biggl[-\frac{1}{{\D_0'}^2}\bigl(\eps-\eps_\mu (1-\q)\bigr)^2\biggr]
                 \nonumber\\
  &\hspace{30pt}\cdot
    \exp\Of[-\frac{1}{\D_0^{\,2}}\Of{a_{\mu i}-\xi-\eta_\nu-\eps_\mu+\eps+b_0\D_0}^2]
    \nonumber\\[2pt]
  &=\frac{1}{\spi\D_2}
  \exp\Of[-\frac{1}{\D_2^{\,2}}
    \Of{a_{\mu i}-\xi-\eta_\nu-\eps_\mu\hspace{.5pt}\q
    +b_0\D_0}^2],
\end{align}
where
$\D_2^{\,2}\equiv{\D_0'}^2+\D_0^{\,2}=4\D_\Q^{\,2}\left(1-\half \q\right)+4\D_\RR^{\,2}$.  Following the same steps as from \eqref{eq:Im_Gtilde} to \eqref{eq:Wieltjes}, the Wieltjes transformation becomes
\begin{align}
  \label{eq:gauss_Wieltjes}
  \tG_{\mu i}^t(\l_n)
  =\frac{\spi}{\D_2}\e^{-\tv^2}\Of{-\erfi(\tv)+\i\hspace{1pt}}
\end{align}
where $\tv\equiv\frac{1}{\D_2}\Of{a_{\mu i}-\l_n-\eta_\nu-\eps_\mu \hspace{.3pt} \q+b_0\D_0}=\frac{1}{\D_2}(x+\eta_\mu-\eta_\nu \allowbreak -\eps_\mu \hspace{.3pt} \q+b_0\D_0)$.

To first order in $t$, we hence obtain
  \begin{align}
    \label{eq:gauss_gamma_mu}
    \g\of(\eps)=t\frac{\spi}{\D_2}
    \e^{-\teps^{\,2}}
  \end{align}
and
  \begin{align}
    \label{eq:gauss_eta_mu}
    \eta\of(\eps)=-t\frac{\spi}{\D_2}
    \e^{-\teps^{\,2}}\erfi\of(\teps),
  \end{align}
where $\teps\equiv\frac{1}{\D_2}\Of{\eps \hspace{1pt} \q-b_0\D_0}$.


%

\end{document}